\documentclass[graybox]{svmult}

\usepackage{mathptmx}       
\usepackage{helvet}         
\usepackage{courier}        
\usepackage{type1cm}        
\usepackage{makeidx}         
\usepackage{graphicx}        
\usepackage{multicol}        
\usepackage[bottom]{footmisc}

\usepackage[authoryear]{natbib}
\usepackage{subfigure}

\makeindex             

\newcommand{\teff}{$T_{\rm{eff}}$}
\newcommand{\msun}{M$_{\odot}$}
\newcommand{\lL}{\ifmmode \log \frac{L}{L_{\odot}} \else $\log \frac{L}{L_{\odot}}$\fi}
\newcommand{\logg}{$\log g$}

\begin{document}

\title*{Empirical properties of Very Massive Stars}
\author{Fabrice Martins}
\institute{Fabrice Martins \at LUPM, Universit\'e Montpellier 2, CNRS, Place Eug\`ene Bataillon, F-34095 Montpellier, France, \email{fabrice.martins@univ-montp2.fr}}
%
%
\maketitle


\abstract{In this first chapter we present the properties of the most massive stars known by the end of 2013. We start with a summary of historical claims for stars with masses in excess of several hundreds, even thousands of solar masses. We then describe how we determine masses for single stars. We focus on the estimates of luminosities and on the related uncertainties. We also highlight the limitations of evolutionary models used to convert luminosities into masses. The most luminous single stars in the Galaxy and the Magellanic Clouds are subsequently presented. The uncertainties on their mass determinations are described. Finally, we present binary stars. After recalling some basics of binary analysis, we present the most massive binary systems and the estimates of their dynamical masses. }

%
\section{Historical background and definition}
\label{hist}

Massive stars are usually defined as stars with masses higher than 8 M$_{\odot}$. In the standard picture of single star evolution, these objects end their lives as core collapse supernovae. Unlike lower mass stars, they go beyond the core carbon burning phase, and produce many of the elements heavier than oxygen. In particular, they are major producers of $\alpha$ elements. Consequently, massive stars are key players for the chemical evolution of the interstellar medium and of galaxies: the fresh material produced in their cores is transported to the surface and subsequently released in the immediate surroundings by powerful stellar winds (and the final supernova explosion). The origin of these winds is rooted in the high luminosity of massive stars (several $10^4$ to a few $10^6$ times the solar luminosity). Photons are easily absorbed by the lines of metals present in the upper layers, which produces a strong radiative acceleration sufficient to overcome gravity and to accelerate significant amounts of material up to speeds of several thousands of km s$^{-1}$ \citep{cak75}. Mass loss rates of $10^{-9}$ to $10^{-4}$ M$_{\odot}$ yr$^{-1}$ are commonly observed in various types of massive stars. Another property typical of massive stars is their high effective temperature. T$_{\rm eff}$ exceeds 25000 K on the main sequence. In the latest phases of evolution, massive stars can be very cool (about 3500 K in the red supergiant phase) but also very hot (100000 K in some Wolf-Rayet stars) depending on the initial mass. A direct consequence of the high temperature is the production of strong ionizing fluxes which create HII regions.

Spectrally, massive stars appear as O and early B (i.e. earlier than B3) stars on the main sequence. Once they evolve, they become supergiants of all sorts: blue supergiants (spectral type O, B and A), yellow supergiants (spectral type F and G) and red supergiants (spectral type K and M). The most massive stars develop very strong winds which produce emission lines in the spectra: these objects are Wolf-Rayet stars. The strong mass loss of WR stars peels them off, unravelling deep layers of chemically enriched material. WN stars correspond to objects showing the products of hydrogen burning (dominated by nitrogen), while WC (and possibly WO) stars have chemical compositions typical of helium burning (where carbon is the main element). For stars more massive than about 25 M$_{\odot}$, there exists a temperature threshold \citep[the Humphreys-Davidson limit, see][]{hd94}) below which stars are expected to become unstable, the ratio of their luminosity to the Eddington luminosity reaching unity. This temperature limit is higher for higher initial masses. Stars close to the Humphreys-Davidson limit are usually Luminous Blue Variable objects (such as the famous $\eta$ Car). 

From the above properties, massive stars are defined as stars with masses higher than 8 M$_{\odot}$. But the question of the upper limit on the mass of stars is not settled. For some time, it was thought that the first stars formed just after the big-bang had masses well in excess of 100 M$_{\odot}$, and likely of 1000 M$_{\odot}$ \citep{bromm99}. The reason was a lack of important molecular cooling channels, favouring a large Jeans mass. Recent advances in the physics of low metallicity star formation have shown that masses of a few tens ``only'' could be obtained if feedback effects are taken into account \citep{hosokawa11}. At the same time, 3D hydrodynamical simulations of massive star formation at solar metallicity have been able to create objects with M $>$ 40 M$_{\odot}$ through accretion \citep{krumholz09}, a process long thought to be inefficient for massive stars because of the strong radiative pressure.

Observationally, the existence of an upper mass limit for stars has always been debated actively. 
The method most often used to tackle this question relies on massive clusters. The idea is to determine the mass function of such clusters and to look for the most massive component. The mass function is extrapolated until there is only one star in the highest mass bin. The mass of this star is the maximum stellar mass expected in the cluster (M$_{max}$). M$_{max}$ is subsequently compared to the mass of the most massive component observed in the cluster (M$_{max}^{obs}$). If M$_{max}^{obs}$ $<$ M$_{max}$ (and if all massive stars in the cluster are young enough not to have exploded as supernovae), then the lack of stars in the mass range  M$_{max}^{obs}$--M$_{max}$ is attributed to an upper mass cut-off in the mass function. \citet{wk04} used this method to infer an upper mass limit of about 150 \msun. Their analysis relied on the young cluster R136. Their conclusions were based on the results of \citet{mh98} who obtained masses of about 140--155 \msun\ for the most massive members. We will see later that the most massive members of R136 may actually be more massive than 150 \msun, which could slightly change the conclusions of Weidner \& Kroupa. Following the same method, \citet{oc05} determined the mass of the most massive member of a cluster as a function of the number of cluster components and of the upper mass cut-off. Using both R136 and a collection of OB associations, they confirmed that an upper mass limit between 120 and 200 \msun\ should exist to explain the maximum masses observed. The studies of \citet{wk04} and \citet{oc05} rely on a statistical sampling of the mass function, without taking into account any physical effects that might alter the formation of massive stars. The subsequent work of \citet{wk06,wkb10} show that a random sampling of the initial mass function may not be the best way of investigating the relation between the cluster mass and the mass of its most massive component. Feedback effects once the first massive stars are formed might be important, stopping the formation of objects in the mass bin M$_{max}^{obs}$--M$_{max}$. This could explain that in the Arches cluster no star with masses in excess of 130 \msun\ is observed while according to \citet{figer05}, there should be 18 of them. Therefore, very massive stars are important not only for stellar physics, but also for star formation and the interplay between stars and the interstellar medium, both locally and on galactic scales. 
 The reminder of this chapter focuses on the search for these objects, and their physical properties.  

\vspace{0.5cm}

Many of the stars with masses claimed to be higher than 100 \msun\ are located in the Magellanic Clouds. The most striking example is certainly that of R136, the core of the 30 Doradus giant HII region in the Large Magellanic Clouc (LMC) where the metallicity is about half the solar metallicity. Using photographic plates, \citet{feitz80} showed that R136 is made of three components (a, b and c) separated by $\sim$ 1''. R136a is the brightest and bluest object. Based on its effective temperature (50000 $<$ T$_{\rm eff}$ $<$ 55000 K) and bolometric luminosity of R136a (3.1 $\times 10^7$ L$_{\odot}$), \citet{feitz80} estimated a lower and upper mass limit of 250 and 1000  M$_{\odot}$ respectively. This made R136a the most massive star at that time. \citet{cas81} obtained an ultraviolet spectrum of R136a with the \textit{International Ultraviolet Explorer} telescope and confirmed the hot and luminous nature of R136a: they obtained a temperature of 60000 K and a luminosity close to $10^8$ L$_{\odot}$. \cite{cas81} compared the morphology of the UV spectrum to that of known early O and WN stars. They concluded that the large terminal velocity deduced from the blueward extension of the P-Cygni profiles (3400 km s$^{-1}$) and the shape of the CIV 1550 line (with a non zero flux in the blue part of the profile) was incompatible with a collection of known massive stars following a standard mass distribution. The authors favoured the solution of a unique object to explain these signatures. This object should have a mass of about 2500 M$_{\odot}$ and a mass loss rate of 10$^{-3.5 \pm\ 1.0}$ M$_{\odot}$ yr$^{-1}$, far in excess of any other known O or Wolf-Rayet star. A similar mass was estimated for the progenitor of the peculiar supernova SN~1961V in the galaxy NGC~1058 \citep{utrobin84}. The width of the maximum emission peak in the light curve of the supernova together with the bright magnitude of the progenitor were only reproduced by hydrodynamical models with masses of the order 2000 M$_{\odot}$.  

The high luminosity is often the first criterion to argue for very massive objects. We will return to this at length in Sect.\ \ref{vms_sing}. Other example of claims for very massive stars based on luminosity in the Magellanic Clouds exist. \citet{humph83} reported on the brightest stars in the Local Group. She listed the brightest blue and red supergiants of six galaxies (Milky Way, SMC, LMC, M33, NGC~6822 and IC~1613). In the Magellanic Clouds, several stars reached bolometric magnitudes of -11 (LMC) and -10 (SMC), corresponding to luminosities in excess of 10$^{6}$ L$_{\odot}$ and thus masses larger than 100 M$_{\odot}$. \citet{kud89} studied the most massive cluster in the SMC, NGC~346, and estimated a mass of 113$^{+40}_{-29}$ M$_{\odot}$ for the brightest component, NGC~346-1.

\begin{figure}[t]
\begin{center}
\includegraphics[width=11cm]{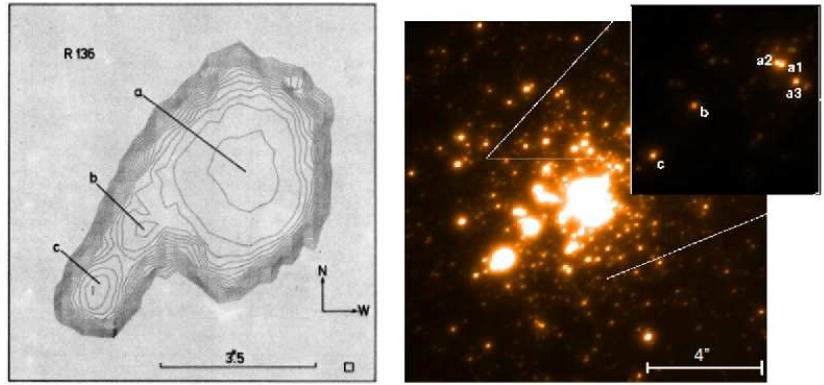}
\caption{R136 cluster in the giant HII region 30 DOradus (Large Magellanic Cloud). \textit{Left panel}: photographic observation of \citet{feitz80}. \textit{Right panel}: Adaptive Optics observations presented in Fig.\ 1 of Crowther et al.\ (2010, MNRAS, 408, 731 "The R136 star cluster hosts several stars whose individual masses greatly exceed the accepted 150Msolar stellar mass limit''). Reproduced with permission.}
\label{fig1}
\end{center}
\end{figure}

Since all of the above examples of very massive stars are located in the Magellanic Clouds, the question of distance and crowding rises. On the good side, distance is rather well constrained, so that luminosity estimates are usually more robust than in the Galaxy. On the other hand, the Magellanic Clouds are much further away than any Galactic region: they are more difficult to resolve. This problem turned out to be a key in the understanding of very massive objects. \citet{weigelt85} used speckle interferometry to re-observe R136. They achieved an spatial resolution of 0.09'' which broke the R136a object into 8 components, all located within 1''. The three brightest members -- R136a1, a2 and a3 -- have similar magnitudes and are separated by a few tenths of arcseconds (Fig.\ \ref{fig1}, right). The 1000 M$_{\odot}$ star in R136a had long lived. Fig.\ \ref{fig1} illustrates the improvements in the imaging capabilities between the study of \citet{feitz80} and the recent paper by \citet{paul10}. Taking advantage of the developments in photometric observations (CCDs, adaptive optics) and image analysis (deconvolution), Heydari-Malayeri et al.\ showed in a series of papers that several of the claimed very massive stars were in fact multiple objects. HDE~268743, one of the bright LMC blue supergiant listed by \citet{humph83}, was first decomposed into six components \citep{hey88} before further observations with AO systems revealed not six but twelve stars. The mass of the most massive objects was estimated to be $\sim$ 50 M$_{\odot}$ (compared to more than 100 M$_{\odot}$ for the single initial object). Similarly, NGC~346-1 turned out to be made of at least three components \citep{hey91}, its mass shrinking from 113 (single object) to 58 M$_{\odot}$ (most massive resolved component). Two additional Magellanic Clouds very bright stars (Sk~157 and HDE~269936) were resolved into at least ten components by \citet{hey89}.

Spatial resolution is thus crucial to understand the nature of very massive stars. But mass estimates also rely on a number of models, for both stellar interiors and stellar atmospheres. The case of the Pistol star in the Galactic Center is an example of the effects of model improvements on mass estimates. \citet{figer98} studied that peculiar object using infrared spectroscopy and atmosphere models. Pistol is a late type massive star, probably of the class of Luminous Blue Variables. Figer et al.\ obtained a temperature of 14000 to 20000 K. Form that, they produces synthetic spectral energy distributions and fitted the observed near infrared photometry. They obtained luminosities between 4 $\times\ 10^{6}$ and 1.5 $\times\ 10^{7}$ L$_{\odot}$, corresponding to masses in the range 200-250 M$_{\odot}$. \citet{paco09} revisited the Pistol star with modern atmosphere models including the effects of metals on the atmospheric structure and synthetic spectra. They revised the temperature (11800 K) and luminosity (1.6 $\times\ 10^{6}$) of the Pistol star, with the consequence of a lower mass estimate: 100 M$_{\odot}$.   

\vspace{0.5cm}

In the following sections of this chapter, we will illustrate how masses of single and binary stars can be determined. We will focus on the methods to constraint the stellar parameters, and especially the luminosity. We will highlight the assumptions of the analysis and raise the main sources of uncertainties.

%
\section{Very Massive Single Stars}
\label{vms_sing}

In this section we will describe how we determine the properties of single massive stars. We will see how an initial mass can be derived from the luminosity, and we will describe the related uncertainties. We will also explain how the present mass of stars can be obtained from the determination of gravity.

%
\subsection{Atmosphere models and determination of stellar parameters}
\label{s_atmo}

Two types of masses are usually determined for massive single stars. The ``evolutionary'' mass is the most often quoted. It is based on the luminosity of the star and its comparison to predictions of evolutionary calculations which provide a direct relation between L and the initial mass of the star. This is done in the Hertzsprung-Russell diagram. An estimate of the effective temperature of the star is required. The second mass is the ``spectroscopic'' mass. It is obtained from the determination of the surface gravity and its radius. From the definition of the gravity $g$, one has: 

\begin{equation}
M = gR^2/G
\end{equation}

\noindent where $G$ is the constant of gravitation. The radius $R$ is usually obtained from the estimate of both the effective temperature and of the luminosity, since by definition:

\begin{equation}
L = 4 \pi\ R^2 \sigma\ T_{eff}^4
\end{equation}

\noindent with $\sigma$ the Stefan-Boltzmann constant. Thus, the spectroscopic mass requires the knowledge of one more fundamental parameter (the surface gravity) compared to the evolutionary mass. For both masses, the determination of the effective temperature and of the luminosity are necessary. 

The most important parameter to constrain is the effective temperature since once it is known the luminosity can be relatively easily determined. Atmosphere models are necessary to estimate T$_{\rm eff}$. They predict the shape of the flux emitted by the star which can be compared to observations, either photometry or spectroscopy. The spectral energy distribution (SED) and lines strength depend sensitively on the effective temperature (and also to a lesser extent on other parameters). Iterations between models and observations allow to find the best models, and consequently the best temperature, to account for the properties of the star. 

For low and intermediate mass stars, the effective temperature can be obtained from optical photometry. Since the peak of the SED is located around the visible wavelength range, a change in \teff\ is mirrored by a change of optical colors \citep[e.g. B-V or V-I,][]{bcp98}. For (very) massive stars, optical colors are almost insensitive to \teff. Their high effective temperature shifts the SED peak to the ultraviolet wavelength range. The visible range is located is in the Rayleigh-Jeans tail of the flux distribution where the slope of the SED barely depends on \teff. In principle, UV colors could be used to constrain the temperature of massive stars. But the UV range is dominated by metallic lines the strength of which depends on several parameters (metallicity, mass loss rate, microturbulence). Consequently, another way has to be found to estimate \teff. The ionization balance is the standard diagnostic: for higher \teff\ the ionization is higher and consequently lines from more ionized elements are stronger. Classically, for O stars, the ratio of HeI to HeII lines is used: HeII line sare stronger (weaker) and HeI lines weaker (stronger) at higher (lower) \teff. Synthetic spectra are compared to observed He lines: if a good match is achieved, the effective temperature used to compute the synthetic spectrum is assigned to the star. He lines are the best temperature indicators between 30000 and 50000 K. Above (below), HeI (HeII) lines disappear. Alternative diagnostics have to be used. In the high temperature range, more relevant for very massive stars, nitrogen lines can replace helium lines \citep{rivero12}. Their behaviour is more complex than helium lines and uncertainties on \teff\ determinations are larger. 

\vspace{0.5cm}     

From above, we see that the determination of the effective temperature of a (very) massive star requires the use of synthetic spectroscopy and atmosphere models. Atmosphere models are meant to reproduce the level populations of all the ionization states of all elements present in the atmosphere of a star, as well as the shape and intensity of the associated radiation field. Their ultimate goal, for spectroscopic analysis of stars, is to predict the flux emitted at the top of the atmosphere so that it can be compared to observational data. An atmosphere model should account for the radial stratification of: the temperature, the density, velocity fields (if present), the specific intensity and opacities. The latter are directly related to level populations. Ideally, such models should be time-dependent and computed in 3D geometry. In practice, the current generation of atmosphere models for massive stars is far from this. The reasons are the following:

\begin{itemize}

\item[$\bullet$] Atmosphere models for massive stars have to be calculated in non-LTE (non Local Thermodynamic Equilibrium). This means that we cannot assume that the flux distribution at each point in the atmosphere is a blackbody. This assumption is not even valid locally. It would be relevant if radiation and matter were coupled only by collisional processes and were at equilibrium. But the very strong radiation field coming from the interior of the star prevents this situation from happening. Radiative processes are much more important than collisional ones. The populations of atomic levels are governed by radiative (de)population and cannot be estimated by the Saha-Boltzmann equation. Instead, the balance between all populating and depopulating routes from and to lower and higher energy levels has to be evaluated. For instance, to estimate the population of the first energy level above the ground state of an element, we have to know the rate at which electrons from the ground level are pushed into the first level, the rate of the inverse transition (from the first level to the ground state) and similarly from all transitions for levels beyond the second level to/from the first level. The computational cost is thus much larger than if the LTE approximation could be applied.  

\item[$\bullet$] Massive stars emit strong stellar winds. Consequently, their atmosphere are extended, with sizes typically between a few tens and up to one thousand times their stellar radius. Spherical effects are important and the assumption of a thin atmosphere  (plane-parallel assumption) cannot be applied. A spherical geometry has to be adopted. In addition, and more importantly, the winds of massive stars are accelerated. Starting from a quasi static situation at the bottom of the atmosphere, material reaches velocities of several thousands of km s$^{-1}$ above ten stellar radii. Doppler shifts are thus induced, which complicates the radiative transfer calculations. A photon emitted at the bottom of the atmosphere can travel freely throughout the entire atmosphere and be absorbed by a Doppler shifted line only in the upper atmosphere. Non local interaction between light and matter are thus possible.  

\item[$\bullet$] For realistic models, as many elements as possible have to be included. This is not only important to predict realistic spectra with numerous lines from elements heavier than hydrogen and helium. It is also crucial to correctly reproduce the physical conditions in the atmosphere. Indeed, having more elements implies additional sources of opacities which can affect the solution of the rate equations and consequently the entire atmosphere structure. The effects of elements heavier than hydrogen and helium on atmosphere models are known as line-blanketing effects. 

\end{itemize}

The combination of these three ingredients makes atmosphere models for massive stars complex. For a reasonable treatment of non-LTE and line-blanketing effects, they are restricted to stationary and 1D computations. There are currently three numerical codes specifically devoted to the study of massive stars: CMFGEN \citep{hm98}, FASTWIND \citep{puls05} and POWR \citep{hamann04}. They all account for the three key ingredients described above. A fourth one \citep[TLUSTY][]{lh03} assumes the plane-parallel configuration and is thus only adapted to spectroscopy in the photosphere of massive stars. All these models are computed for a given set of input parameters. The effective temperature, the luminosity (or the stellar radius), the surface gravity, the chemical composition and the wind parameters (mass loss rate and terminal velocity) are the main ones. The computation of the atmospheric structure is performed for this set of parameters. Once obtained, a formal solution of the radiative transfer equation is usually done to produce the emergent spectrum. It is this spectrum that is subsequently compared to observations. If a good match is obtained, the parameters of the models are considered to be the physical parameters of the star. CMFGEN and POWR are better suited to the analysis of VMS since as we will see below, VMS usually appear as Wolf-Rayet stars with numerous emission lines from H, He but also metals.

\begin{figure*}[t]
     \centering
\subfigure[]{ \includegraphics[width=.47\textwidth]{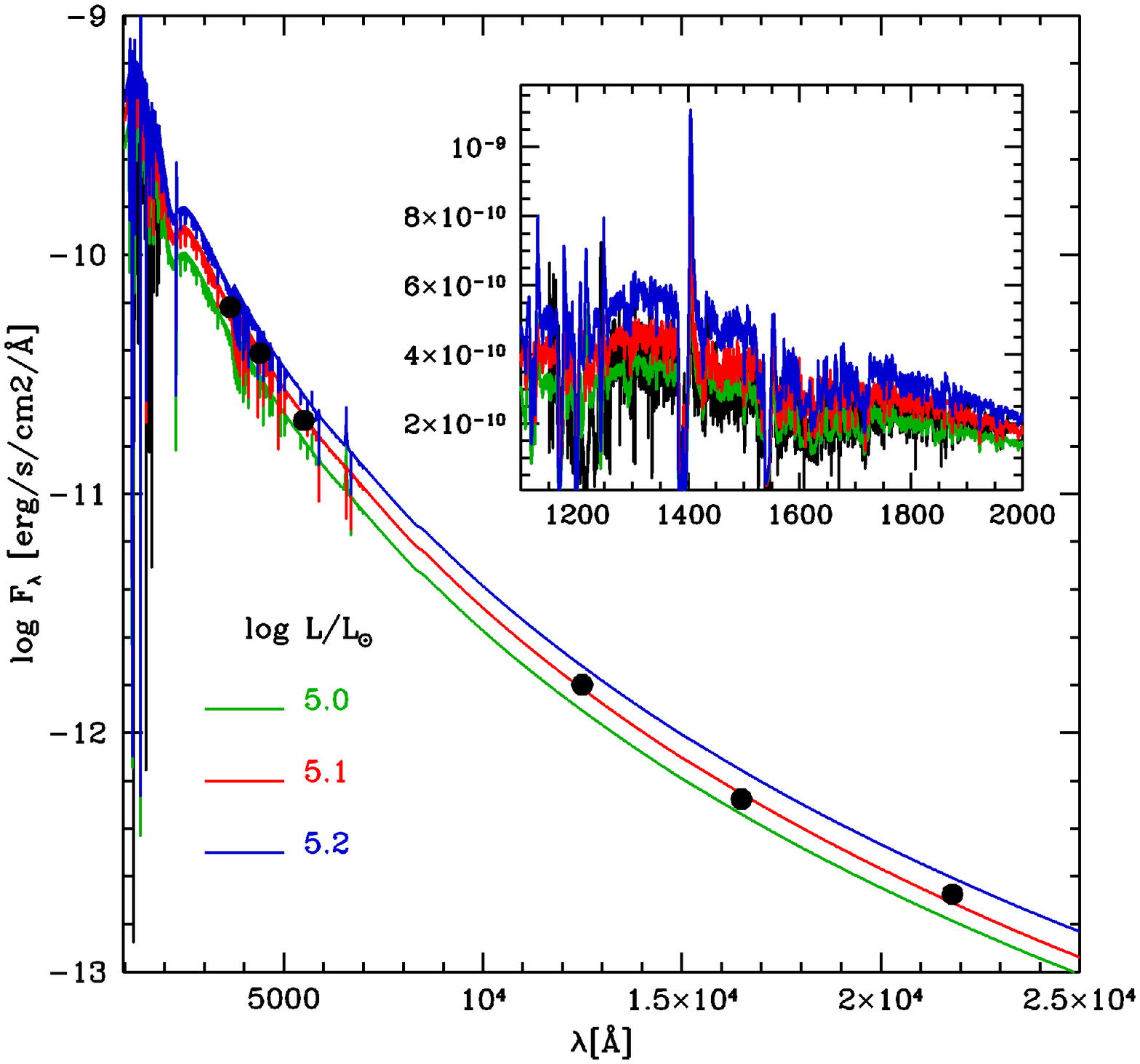}}
     \hspace{0.1cm}
     \subfigure[]{
           \includegraphics[width=.47\textwidth]{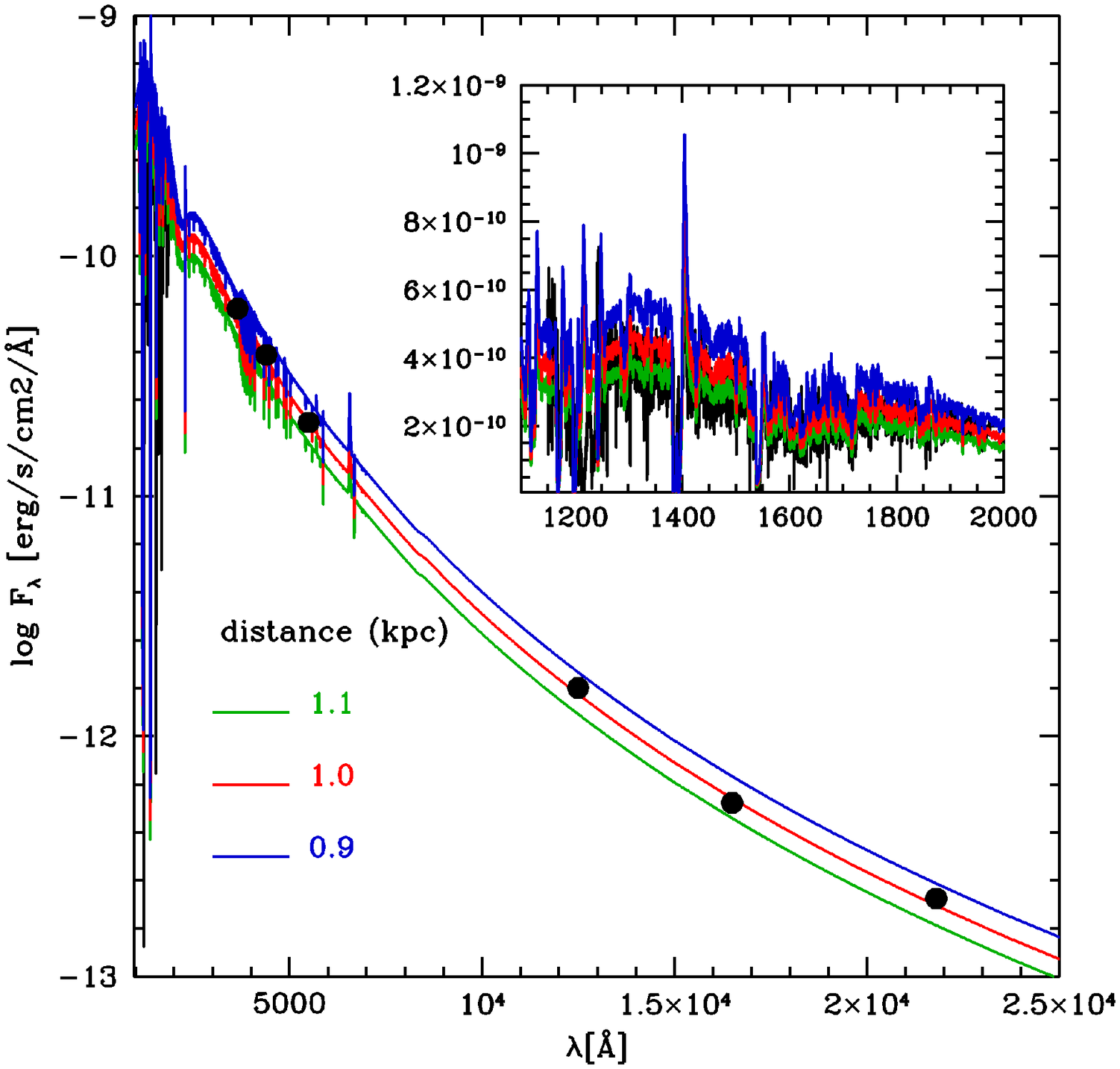}}
     \caption{Determination of the luminosity from the spectral energy distribution. The black line and dots are data for star HD~188209. \textit{Left panel}: the colored lines are three models with luminosities differing by 0.1 dex. \textit{Right panel}: the colored lines are the same model for three different distances. From Martins et al.\ (in prep).}
     \label{fig_L}
\end{figure*}

As explained above, the effective temperature is the first parameter to constrain. Once it is obtained, the luminosity determination can be made. There are usually two ways to proceed: either spectro-photometric data covering a large wavelength region exist and the SED can be fitted, or only photometry in a narrow wavelength region is available, and bolometric corrections have to be used. SED fitting is illustrated in the left panel of Fig.\ \ref{fig_L}. Atmosphere models provide the spectral energy distribution at the stellar surface for the set of input parameters. This flux is scaled by the ratio (R/d)$^2$ where R is the stellar radius and d the distance of the target star. The resulting flux is then compared to the observational spectro-photometric data. In Fig.\ \ref{fig_L} we see the effect of a change of 0.1 dex on the luminosity. Optical and infrared photometry has been used together with flux calibrated UV spectra to build the observed SED. The red model, corresponding to \lL\ = 5.2, best reproduces these data. 
The right panel of Fig.\ \ref{fig_L} illustrates an important limitation of the determination of luminosities for massive stars: the knowledge of distances and their uncertainties. An error of only 10\% on the distance is equivalent to an error of about 0.1 dex on the luminosity. Distance is usually the main contributor to the uncertainty on the luminosity of Galactic objects. 

If a sufficient number of spectro-photometric data is available, SED fitting can be performed rather safely (see Sect.\ \ref{s_L} for limitations). Often, only photometry in the optical or the infrared range can be obtained. This is the case for stars located behind large amounts of extinction. In that case, luminosity has to be determined differently. This is done through an estimate of the bolometric correction. In the following, we will assume that only K-band photometry is available. The first step is to estimate the absolute K band magnitude

\begin{equation}
K = mK - A(K) - DM
\label{eq3}
\end{equation}

\noindent where $mK$ is the observed magnitude, $A(K)$ the amount of extinction in the K band and $DM$ the distance modulus (DM = 5$\times$log(d)-5, d being the distance). As before, a good knowledge of the distance is required. We see that an estimate of the extinction is necessary too. We will come back to this issue in Sect.\ \ref{s_L}. In the second step, we need to add to the absolute K-band magnitude a correction to take into account the fact that we observe only a small fraction of the entire flux. This bolometric correction is computed from atmosphere models, and calibrated against effective temperature. For instance, \cite{mp06} give

\begin{equation}
BC(K) = 28.80 - 7.24 \times\ log(T_{\rm eff})
\label{eq4}
\end{equation} 

\noindent The K-band bolometric correction is thus by definition the total bolometric magnitude minus the K-band absolute magnitude. Said differently, with K and BC(K), we have the total bolometric magnitude and thus the luminosity of the star: 

\begin{equation}
\lL\ = -0.4 \times\ (K + BC(K) - M_{\odot}^{bol})
\label{eq5}
\end{equation} 

\noindent where $M_{\odot}^{bol}$ is the bolometric magnitude of the Sun. Both methods (SED fitting or bolometric corrections) rely on atmosphere models and thus depend on the assumptions they are built on.

\begin{figure}[t]
\begin{center}
\includegraphics[width=7cm]{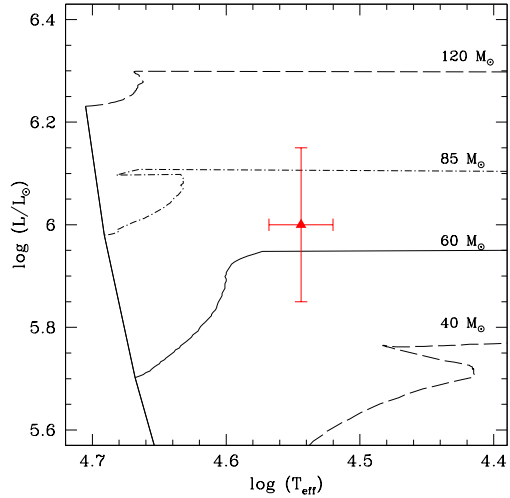}
\caption{Hertzsprung-Russell diagram illustrating the determination of evolutionary masses. Evolutionary tracks are from \citet{mm03}. The red square is for a star with \teff\ = 35000 $\pm$ 2000 K and \lL\ = 6.0 $\pm$ 0.15. The derived initial mass is 68$^{+25}_{-14}$ \msun, while the estimated present-day mass is 50$^{+18}_{-9}$ \msun.}
\label{fig_hr_L}
\end{center}
\end{figure}

\vspace{0.5cm}

Once the effective temperature and luminosity are constrained, the evolutionary mass can be determined.
Fig.\ \ref{fig_hr_L} shows a classical Hertzsprung-Russell diagram with the position of a bright O supergiant shown by the red symbol. From interpolation between the evolutionary tracks, we can estimate a present-day mass of 50$^{+18}_{-9}$ \msun\ and an initial mass of 68$^{+25}_{-14}$ \msun. The difference between both estimates is due to mass loss through stellar winds. These masses are the evolutionary masses introduced at the beginning of this section. They depend on the way the interpolation between tracks is done and most importantly on the tracks themselves (Sect.\ \ref{s_tracks}). From Fig.\ \ref{fig_hr_L}, we see that very massive stars are objects with luminosities larger than one million times the Sun's luminosity.

The second mass estimate that can be given for single massive stars is the spectroscopic mass. As explained above, it is obtained from the surface gravity and an estimate of the stellar radius. The surface gravity, \logg, is obtained from the fit of Balmer, Paschen or Brackett lines in the optical/infrared range. Their width is sensitive to pressure broadening, especially Stark broadening. Stark broadening corresponds to a perturbation of the energy levels due to the electric field created by neighbouring charged particles. Broadening is thus stronger in denser environments, and consequently in stars with larger surface gravity. Hence, the width and strength of lines sensitive to Stark broadening effects are good estimates of \logg. The most commonly used spectral diagnostics of surface gravity are H$\beta$ and H$\gamma$ in the visible range. In the infrared, Br$\gamma$ and Br10 are the best indicators. Synthetic spectra computed for a given \logg\ are directly compared to the observed line profile. An accuracy of 0.1 dex on \logg\ is usually achieved. This corresponds to an uncertainty of about 25\% on the stellar mass (without taking into account any error on the stellar radius). 

The spectroscopic mass is more difficult to obtain than the evolutionary mass since it requires the observation of photospheric hydrogen lines. Such lines are sensitive to mass loss rate. When the wind strength becomes large, emission starts to fill the underlying photospheric profile leaving usually a pure emission profile, preventing the determination of \logg. Unfortunately, this is often the case for very massive stars (see Sect.\ \ref{s_candidates_sing}) which are very luminous and consequently have strong stellar winds. 

The evolutionary mass and the spectroscopic mass should be consistent. However, as first pointed out by \citet{her92}, the former are often systematically larger than the latter. Improvements in both stellar evolution and atmosphere models have reduced this discrepancy \citep[e.g.][]{mokiem07,jc13}, but the problem is still present for a number of stars. At present, the reason(s) for this difference is (are) not clear. Studies of binary systems tend to indicate that the evolutionary masses and the dynamical masses obtained from orbital solutions (see Sect.\ \ref{vms_bin}) are in good agreement below 30 to 50 \msun. Above, that limit, no clear conclusion can be drawn \citep{burk97,wv10,massey12}. 
The general conclusion is that there are at least two types of mass estimates for massive stars and that currently, no preference should be given to any of them.

\vspace{0.5cm}

In this section, we have presented the mass determinations for massive stars. For very massive objects, the evolutionary masses are usually quoted because of the shortcomings of the surface gravity determination. Evolutionary masses rely heavily on luminosity estimates and on the relation between luminosity and mass. In the next two sections, we will present the uncertainties related to both of them.

%
\subsection{Uncertainties on the luminosity}
\label{s_L}

We now focus on the errors that enter the determination of the stellar luminosity. We assume we are dealing with single stars. In case of multiplicity, the determinations of L are obviously overestimated by an amount which depends on the number of companions and their relative brightness. 

We have seen above that the luminosity of a star could be obtained from the monochromatic magnitude, a bolometric correction, an estimate of the extinction and of the distance (see Eqs.\ \ref{eq3} to \ref{eq5}). This method is useful when not enough data are available to fit the entire SED. This is often the case for objects hidden behind large amounts of extinction. We have used this set of equations to compute the effects of uncertainties of several quantities on the derived luminosity. The results are summarized in Table \ref{tab_erL}. We have considered the case of a star with \lL\ $\sim$ 6.0. We have assumed it was observed in the K band. The distance and extinction are consistent with a position in the Galactic Center, a place where many massive stars are found. We have assumed typical errors on the magnitude, extinction, distance and effective temperature. The latter directly affects the uncertainty on the bolometric correction (Eq. \ref{eq4}). From Table \ref{tab_erL}, we see that the largest error budget is due to the uncertainties on the effective temperature and distance when the latter is poorly constrained. When combining all the sources of uncertainty, we get an error in the luminosity of about 0.17 dex (for a 10\% uncertainty on the distance). 

In Table \ref{tab_erL}, we also provide estimates of the variations in mass estimates due to the above uncertainties. The evolutionary tracks of \citet{mm03} have been used in this test case. The parameters we have chosen are those of a star with an initial mass of about 78 \msun\ according to the Meynet \& Maeder tracks. The individual errors induce changes of the initial mass between 7 and 18 \msun. The combined effects correspond to an uncertainty of 28 \msun, or 35\%. If the distance is poorly known (e.g. 25\% error, see last line of Table \ref{tab_erL}), the uncertainty can even reach almost 50\%.

\begin{table}[t]
\begin{center}
\caption{Effect of various uncertainties on luminosity and mass estimate$^a$. }
\label{tab_erL}       
%
%
\begin{tabular}{p{4cm}p{2cm}p{2cm}p{2cm}}
\hline\noalign{\smallskip}
Error            & $\Delta$ \lL\ & $\Delta$ M$_{init}$ [\msun] & $\Delta$ M/M [\%] \\
\noalign{\smallskip}\svhline\noalign{\smallskip}
$\Delta$ m = 0.1       &   0.04        &     7   & 9.0  \\
$\Delta$ A = 0.2       &   0.08        &    13   & 16.7  \\
$\Delta$ d =0.1        &   0.09        &    15   & 19.2  \\
$\Delta$ \teff = 3000  &   0.11        &    18   & 23.1  \\
Combination of above errors &   0.17   &    28   & 35.9  \\
$\Delta$ d =0.25       &   0.22        &    36   & 46.1  \\
\noalign{\smallskip}\hline\noalign{\smallskip}
\end{tabular}
\end{center}
$^a$ Calculations are for a star at 8kpc, behind an extinction of 3 magnitude, with an observed magnitude of 11.2 and an effective temperature of 35000 K. Photometry is taken in the K band. The luminosity and evolutionary mass of such a star would be \lL\ = 6.06 and 78 \msun\ (initial mass). The mass is estimated using the evolutionary tracks of \citet{mm03}. 
\end{table}

\begin{figure*}[t]
     \centering
\subfigure[]{ \includegraphics[width=.47\textwidth]{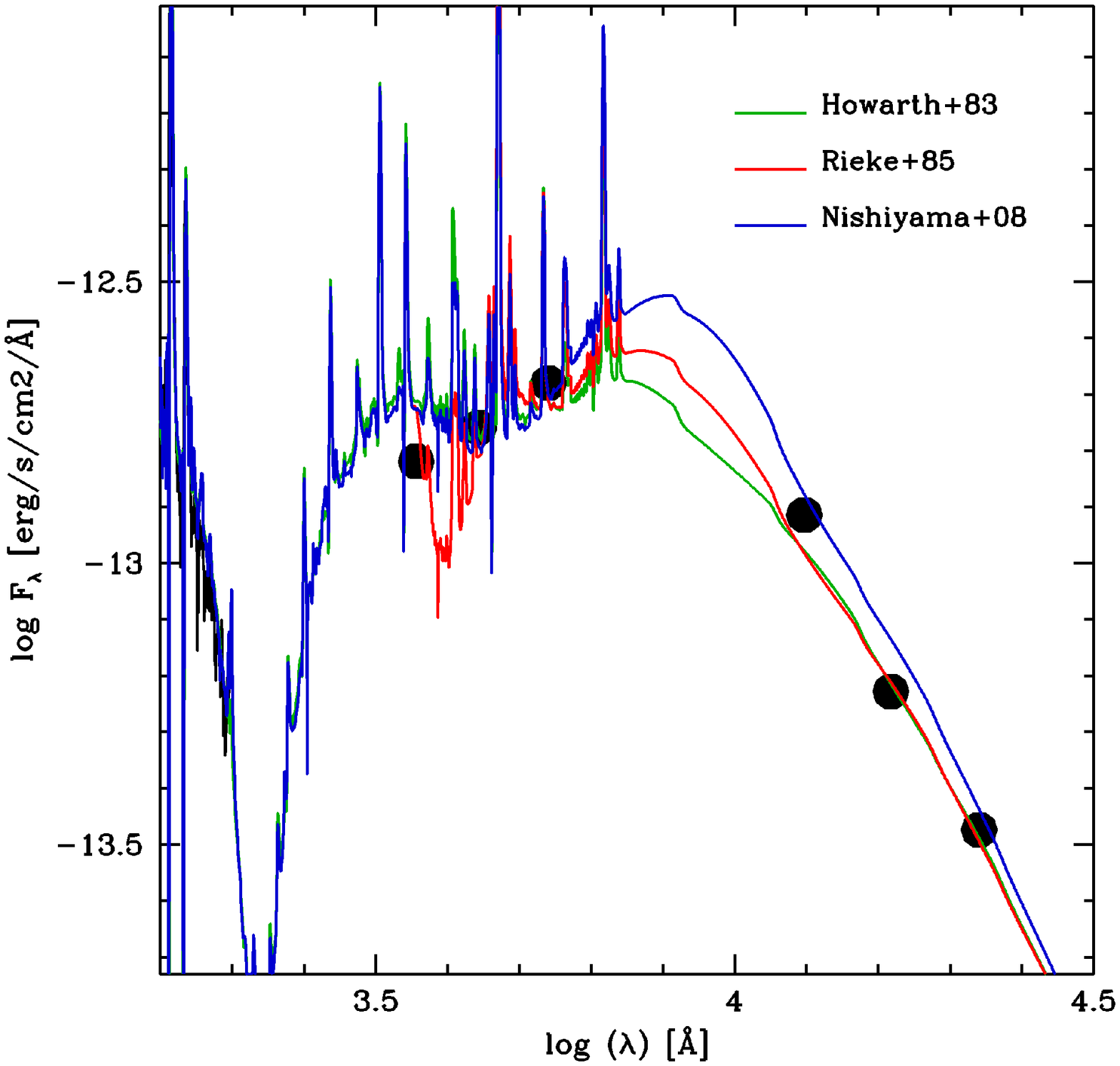}}
     \hspace{0.1cm}
     \subfigure[]{
           \includegraphics[width=.47\textwidth]{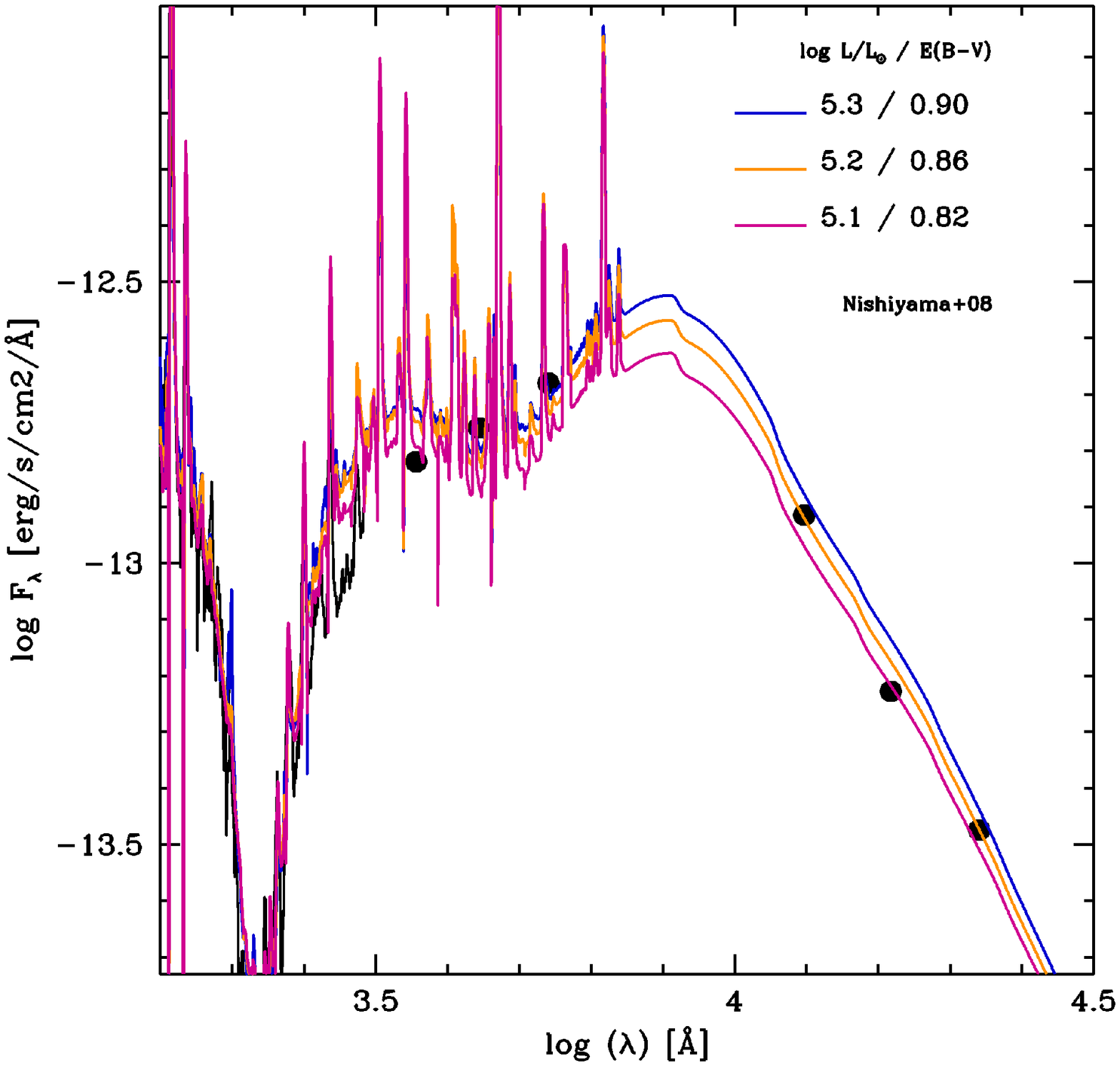}}
     \caption{Extinction law and luminosity determination. \textit{Left panel}: Effect of various extinction law on the luminosity determination. The data are for the Galactic star WR~18 (black line and dots). The colored lines are models redenned with different extinction laws in the optical/infrared range. In the UV, the extinction law is that of \citet{seaton79}.  \textit{Right panel}: illustration of the luminosity tuning necessary to fit the SED with the extinction law of \citet{nishi09}.}
     \label{fig_extlaw}
\end{figure*}

Another source of uncertainty not taken into account in the above estimates is the shape of the extinction law. In Table \ref{tab_erL} we have assumed a K-band extinction of 3.0. However, depending on the extinction law, the stellar flux will be redenned differently and for the same star, different values of the extinction can be obtained. As a consequence, the luminosity estimate will be affected. An illustration of this effect is given in the left panel of Fig.\ \ref{fig_extlaw}. Observational data (in black) for the Galactic star WR~18 are compared to a model with \lL\ = 5.3 (colored lines). In the UV, the extinction of \citet{seaton79} is adopted. In the optical/infrared, three different extinction laws have been used: \citet{howarth83} (green), \citet{rieke85} (red) and \citet{nishi09} (blue). All models assume E(B-V)=0.9 and R$_{V}$=3.2. The UV part of the SED is correctly reproduced, except in the region around 3000 \AA. The optical and infrared flux are different depending on the extinction curve. The laws of \citet{howarth83} and \citet{rieke85} are relatively similar in the infrared but differ in the optical. The Nishiyama et al.\ law leads to a larger flux in the infrared compared to the other too laws. If we were to adopt the Nishiyama et al.\ law, we would need to reduce the luminosity to reproduce the infrared part of the SED. This is shown in the right panel of Fig.\ \ref{fig_extlaw}. The model with \lL\ = 5.3 and E(B-V) = 0.9 is shown in blue. Two additional models are shown: one with \lL\ = 5.2 and E(B-V)=0.86 (orange) and one with \lL\ = 5.1 and E(B-V)=0.82 (magenta). A luminosity intermediate between that of the two new models better reproduces the SED. Hence, using the Nishiyama et al.\ extinction leads to a downward revision of the luminosity by $\sim$ 0.15 dex. At luminosities of $\sim$2 $\times 10^5$ L$_{\odot}$, such a change corresponds to a reduction of the initial mass by 4 \msun, or 12\%. 

In Table \ref{tab_erL} we finally show the influence of the uncertainty on the effective temperature. It affects the bolometric correction and consequently the luminosity. A typical error of 3000 K on \teff\ (usually found for infrared studies) corresponds to a change in luminosity by 0.11 dex, and thus to an uncertainty of about 20-25\% on the initial mass. This estimate does not take into account the uncertainty in the relation between bolometric correction and temperature. It only accounts for the effect of \teff\ on BC for the relation given in Eq.\ \ref{eq4}. Different model atmospheres provide slightly different calibrations of bolometric corrections. This adds another source of error in the mass estimate. 

\vspace{0.5cm}

In conclusion, various uncertainties in the quantities involved in the luminosity determination lead to a typical error of about 0.15--0.20 dex on \lL. If the distance is poorly known, this uncertainty is larger. This translates into error on the estimate of the initial mass of the order 10 to 50\%.

%
\subsection{Uncertainties in evolutionary tracks}
\label{s_tracks}

In the previous section, we have seen how the luminosity determination was affected by uncertainties in various observational quantities. We now focus on the uncertainties involved in the interpretation of the determined luminosity. As explained previously, the determination of the initial or present mass relies on comparison with evolutionary tracks (see Sect.\ \ref{s_atmo}). Such models rely on different assumptions to take into account the physical processes of stellar evolution. Consequently, they produce different outputs. In the following we will compare the public tracks of \citet{brott11} and \citet{ek12}. We refer to \citet{mp13} for a detailed comparison of various tracks. 

\begin{figure}[t]
\begin{center}
\includegraphics[width=7cm]{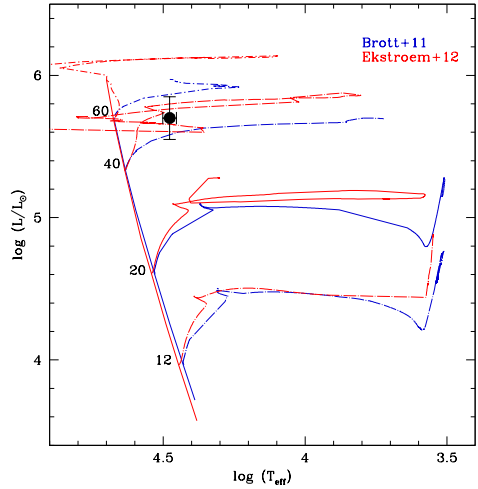}
\caption{Hertzsprung-Russell diagram with the evolutionary tracks of \citet{brott11} (blue) and \citet{ek12} (red). The Brott et al.\ tracks are for an initial rotational velocity of 300 km~s$^{-1}$ while those of Ekstroem et al.\ assume a ratio of initial to critical rotation of 0.4. }
\label{fig_gebr}
\end{center}
\end{figure}

Fig.\ \ref{fig_gebr} shows the evolutionary tracks for Galactic stars from these two public grids of models. The tracks from \citet{brott11} are only available up to 60 \msun, so we do not show the higher mass models of \citet{ek12}. There are many differences between both sets of tracks \citep[see][]{mp13}. The most important one for the sake of mass determination is the very different luminosities for a given mass. Looking at the 40 \msun\ tracks, we see that the Ekstroem et al.\ tracks are about 0.2 dex more luminous than the Brott et al.\ tracks beyond the main sequence. This offset is smaller at lower masses, and larger at higher masses. The direct consequence is that a lower initial mass is needed by the Ekstroem et al.\ tracks to reproduce the observed luminosity of a star. The black dot in Fig.\ \ref{fig_gebr} is an artificial star with \teff\ = 30000 K and \lL\ = 5.7. Using the Ekstroem et al.\ tracks, one would find an initial mass of 37 \msun. For the Brott et al.\ tracks, the initial mass would be 48 \msun. The difference is of the order of 25\%.

The origin for the differences between the predictions of evolutionary tracks are manifold. One of the key effect known to affect the luminosity of a star is the amount of overshooting. The size of the convective core of massive stars is usually defined by the Schwarzschild criterion. But when this criterion applies to the velocity gradient of the convective region, and not to the velocity itself. It means that before reaching a zero velocity, the material transported by convection in the core travels a certain distance above the radius defined by the Schwarzschild criterion. This distance is not well known and is usually calibrated as a function of the local pressure scale height. The effect of this overshooting region is to bring to the core fresh material. This material is burnt and thus contributes to the luminosity of the star. Consequently, the larger the overshooting, the higher the luminosity of the star. 

Another cause of luminosity increase is the effect of rotation. The physical reason is the same as for overshooting. When a star rotates, mixing of material takes place inside the star. Consequently, there is more material available for burning in the stellar core compared to a non-rotating star. The effect is an increase of the luminosity \citep{meym00}. Quantitatively, the increase is between 0.1 and 0.3 dex depending on the initial mass for a moderate rotation of 300 km s$^{-1}$ on the main sequence. Here again, the consequence for mass determinations is that lower masses are determined when tracks with rotation are used. In the case of the tracks of \citet{ek12}, and for the same example as above, a mass of 49 \msun\ would be derived without rotation, compared to 37 \msun\ with rotation. For very massive stars, the effects of rotation might be limited. The reason is the large mass loss rates at very high luminosities, causing efficient braking. However, the effects of rotation have not (yet) been investigated in the very high mass range.

Other parameters affecting the shape of evolutionary tracks are mass loss rate and metallicity. The former impact the evolution of the stars by removing material through stellar winds. Depending on the strength of the winds, the mass of a star at a given time will be different. Since mass and luminosity are directly related, a star with a strong mass loss will have a lower luminosity than a star with a lower mass loss \citep{mm94}. Mass loss rates used in evolutionary calculations come from various sources. Some are empirical, some are theoretical. There are uncertainties associated with mass loss rates, but they are not straightforward to quantify. Clumping is known to affect the mass loss rate determinations in O-type stars \citep{jc05,ful06} but a good handle of its properties is missing. For cool massive stars, there is a wide spread in the mass loss rates of red supergiants \citep{mj11}. Stellar winds are also weaker at lower metallicity (at least for hot massive stars). Their dependence is rather well constrained in the range 0.5 $<$ Z/Z$_{\odot} <$ 1.0, but beyond, mass loss rates are based on extrapolations. Hence, evolutionary calculations, which adopt general (empirical or theoretical) prescriptions, can be adapted to explain the averaged properties of massive stars, but may fail to explain individual objects. They should thus be used with care.

%
\subsection{The best cases for very massive single stars}
\label{s_candidates_sing}

After raising the sources of uncertainty in the determination of the mass of the most massive stars, we now turn to the presentation of the best cases. Since the stellar initial mass function (IMF) is a power law of the mass (at least above $\sim$ 1 \msun), massive stars are very rare. Consequently, we expect to find them more predominantly in clusters or association hosting a large number of stars. Adopting a standard Salpeter IMF, a cluster should have a mass in excess of a few 10$^3$ \msun\ in order to host at least one star with mass in excess of 100 \msun. Thus, very massive stars have to be searched in massive clusters. In addition, massive stars live only a few million years, typically 2 to 3 Myr for objects above 100 \msun\ \citep[e.g.][]{yusof13}. Hence, VMS can only be found in young massive clusters. The best places to look for them would be young super star clusters. Observed in various types of galaxies, these objects have estimated masses in excess of 10$^4$ \msun, some of them reaching several 10$^5$~\msun\ \citep{mengel02,bastian06}. However, none of these super star cluster is known in the local group, preventing current generation of telescopes and instruments to resolve their components individually. 

In the Galaxy and its immediate vicinity, where individual stars can be observed, the best place to look for VMS is thus in young massive clusters. Although there has been a lot of improvement in the last decade in the detection of such objects \citep{figer06,chene13}, only a few are massive and young enough to be able to host VMS. In the Galaxy, the Arches and NGC3603 clusters are so far the two best candidates. In the Magellanic Clouds, R136 in the giant HII region 30~Doradus (see Sect.\ \ref{hist}) is another interesting cluster. Beyond these three cases, known clusters are either too old or not massive enough. In the following, we will describe the evidence for the presence of VMS in the Arches, NGC3603 and R136 clusters. We will also highlight a couple of presumably isolated stars with large luminosities.

%
\subsubsection{The Arches cluster}
\label{s_arches}
 
The Arches cluster is located in the center of the Galaxy. It was discovered in the late 90's through infrared imaging \citep{cotera96,figer99}. First thought to have a top heavy IMF \citep{figer02} it is now considered to host a classical mass function \citep{espinoza09}. It hosts 13 Wolf-Rayet stars and several tens of O stars. All the Wolf-Rayet stars are of spectral WN7-9h. They are very luminous and their properties are consistent with those core-H burning objects \citep{martins08}. Their luminosities, estimated using Eq.\ \ref{eq3} and \ref{eq5}, is larger than 10$^6$ L$_{\odot}$, indicating that they are more massive than 80 \msun. \citet{martins08} adopted a short distance (7.6 kpc) and a low extinction (A$_{K}$=2.8) to obtain these luminosities. Revisiting the properties of two of the Arches WNh stars with a larger distance and extinction, \citet{paul10} derived luminosities larger by about 0.25 dex. Consequently, they reported masses in excess of 160 \msun. This illustrates the importance of extinction and distance in the estimates of the initial mass of VMS, as described in Sect.\ \ref{s_L}.

\begin{figure*}[t]
     \centering
     \subfigure[]{ \includegraphics[width=.47\textwidth]{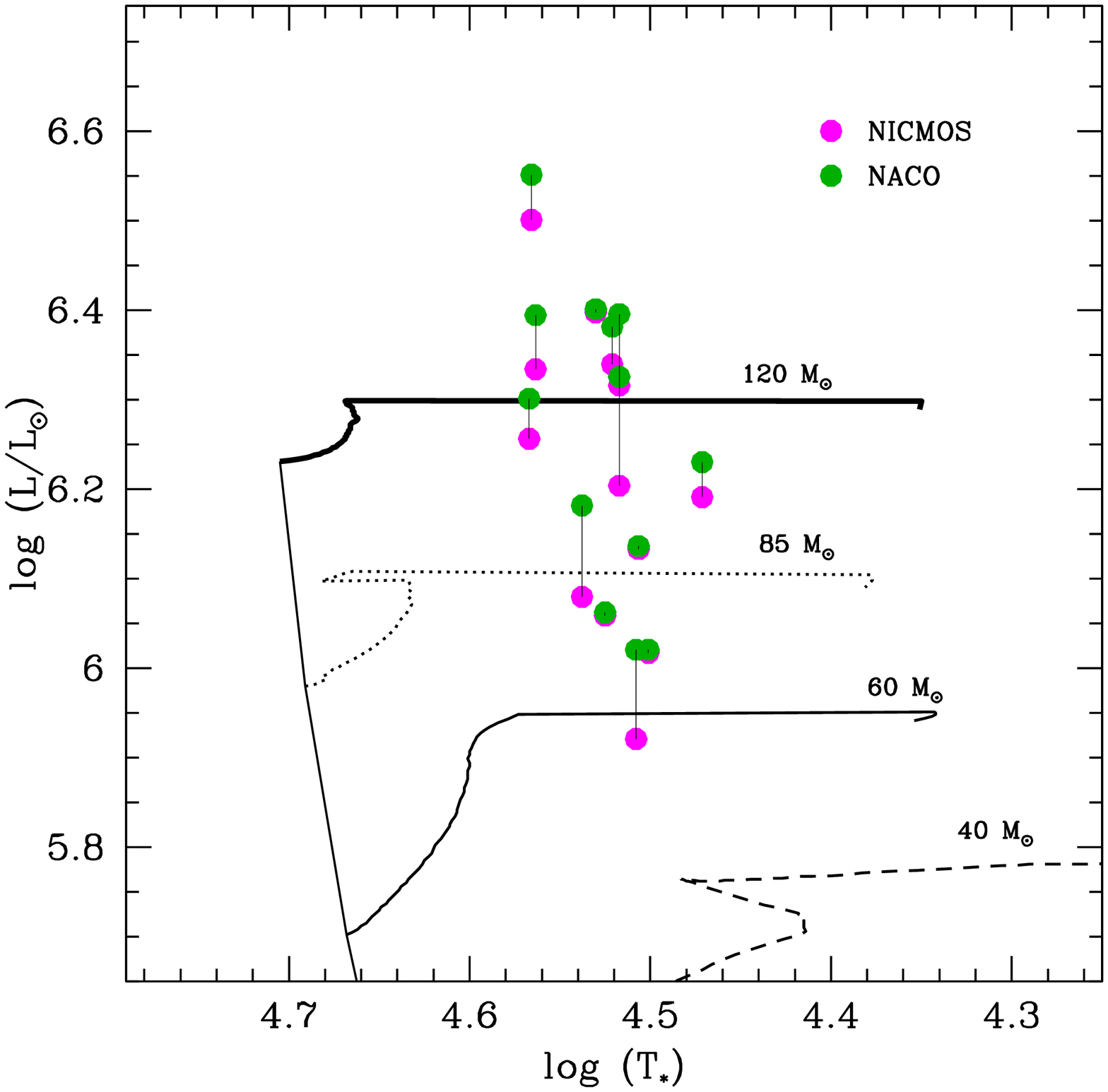}}
     \hspace{0.1cm}
     \subfigure[]{\includegraphics[width=.47\textwidth]{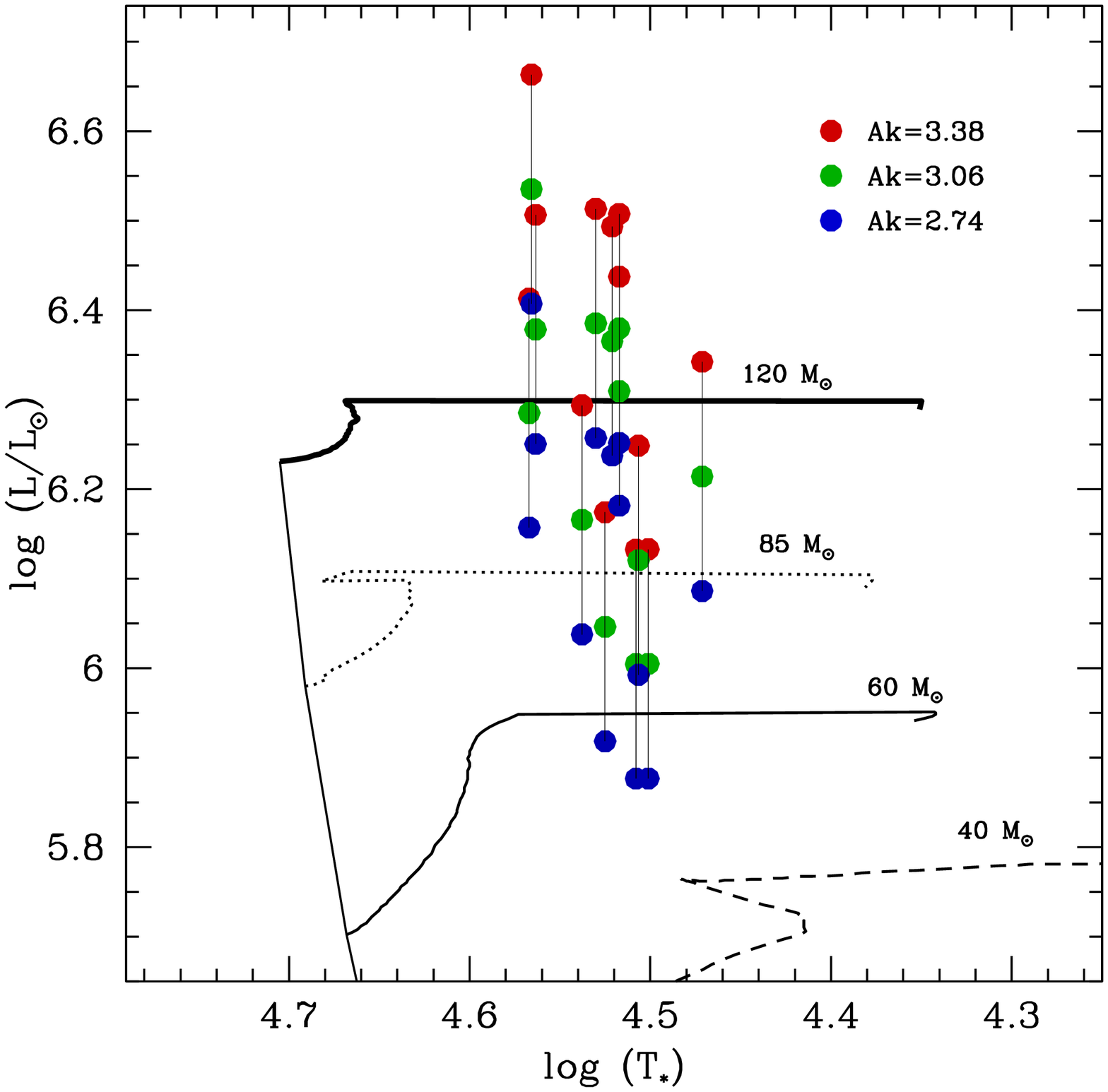}}
     \subfigure[]{\includegraphics[width=.47\textwidth]{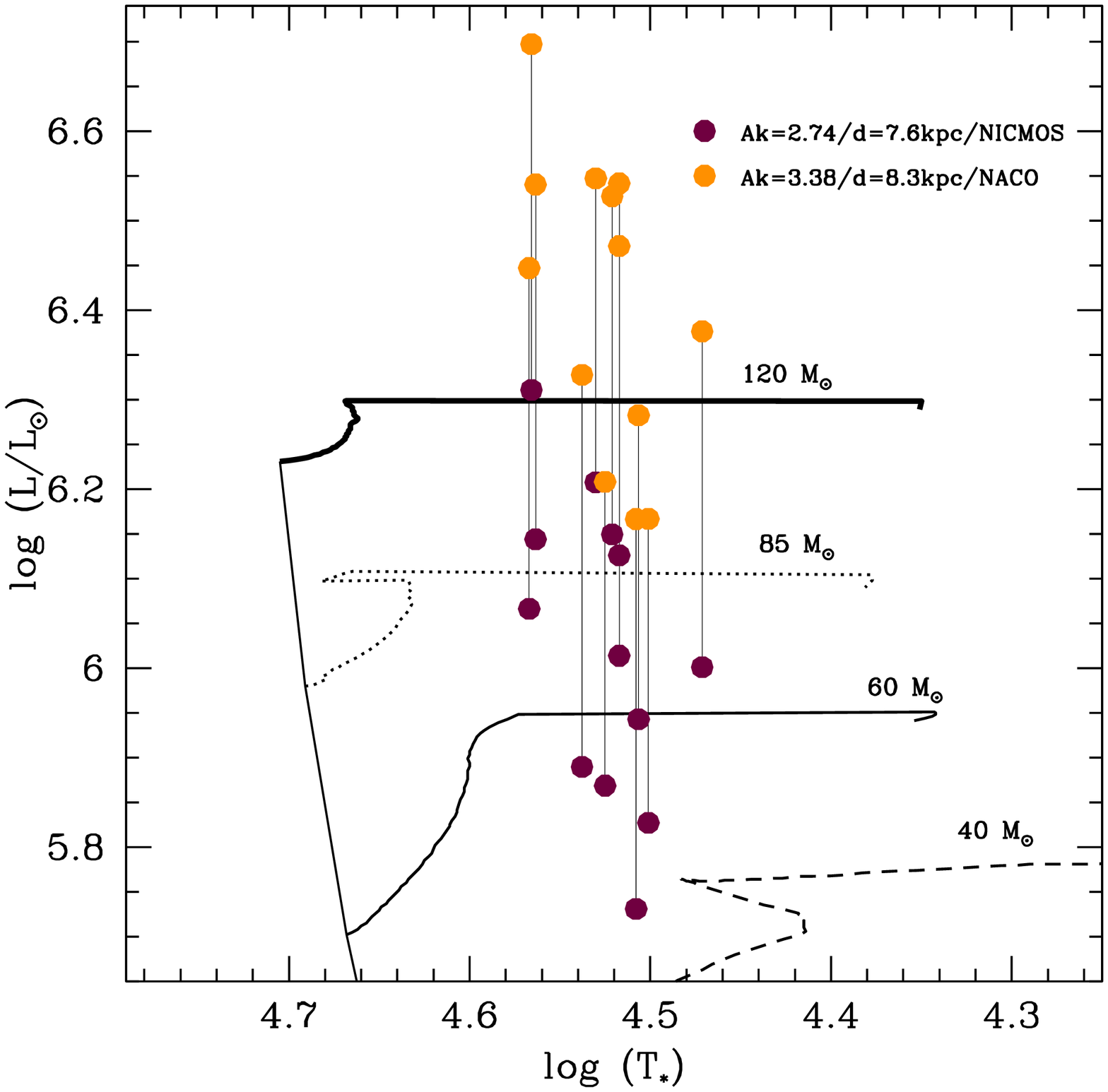}}
     \caption{HR diagram of the Arches cluster. The data points correspond to the WNh stars. The evolutionary tracks are from \citet{mm03} and have Z=0.02. \textit{Upper left panel}: effect of photometry. The luminosity of the stars is computed using both the HST/NICMOS and VLT/NACO photometry \citep{figer99,espinoza09}. The distance is set to 8.0 kpc and effective temperatures are from \citet{martins08}. \textit{Upper right panel}: effect of extinction. An average value of E(H-K)=1.9 is adopted and is transformed into A$_{K}$ using the relations of \citet{rieke85}, \citet{nishi09} and \citet{espinoza09}. \textit{Bottom panel}: extreme values of the Arches luminosities obtained by combining the faintest (brightest) photometry \citep{figer99,espinoza09}, lowest (highest) extinction \citet{rieke85,nishi09} and shortest (largest) distances \citep{eisenhau05,gil09}.}
     \label{fig_arches}
\end{figure*}

\begin{table}[t]
\begin{center}
\caption{Effects of photometry, extinction and distance uncertainty on the initial mass estimate$^a$ for star F12 in the Arches cluster.}
\label{tab_arches}       
%
%
\begin{tabular}{p{4cm}p{2cm}p{2cm}}
\hline\noalign{\smallskip}
mK / A$_{K}$ / d(kpc)      & M$_{init}^a$ [\msun] \\
\noalign{\smallskip}\svhline\noalign{\smallskip}
10.99$^b$ / 3.1$^c$ / 8.0   &   111  \\
10.88$^d$ / 3.1$^c$ / 8.0   &   120  \\
10.88$^d$ / 3.06$^d$ / 8.0  &   117  \\
10.88$^d$ / 3.38$^d$ / 8.0  &   138  \\
10.88$^d$ / 2.74$^d$ / 8.0  &   93   \\
10.99 / 2.74 / 7.6$^e$      &   78   \\
10.88 / 3.38 / 8.3$^e$      &   146  \\
\noalign{\smallskip}\hline\noalign{\smallskip}
\end{tabular}
\end{center}
$^a$ The mass is estimated using the evolutionary tracks of \citet{mm03}. \\
$^b$ Photometry is from \citet{figer99} and \citet{espinoza09}. \\
$^c$ Extinction is from \citet{stolte02}. \\
$^d$ Extinction is computed assuming E(H-K)1.9 and using the redenning laws of \citet{rieke85}, \citet{espinoza09} and \citet{nishi09}. \\
$^e$ Distances are from \citet{eisenhau05} and \citet{gil09}.
\end{table}

In Fig.\ \ref{fig_arches} we present new HR diagrams of the Arches WNh stars. Table \ref{tab_arches} provides the initial mass estimates depending on various assumptions. In this table, we have selected star F12 of \citet{figer02} as a test case. F12 is the hottest object in the HR diagram. Fig.\ \ref{fig_arches} and Table \ref{tab_arches} are meant to further illustrate the role of various observational parameters on mass determinations. The upper left panel shows the effect of photometry. A change of 0.1 magnitude in the K-band photometry results in a change of about 10\% in the initial mass. The effect of extinction is similar. In our estimates, we have assumed a constant value of E(H-K) for all stars. Depending on the extinction law adopted, the ratio of selective to total absorption in the infrared (R$_{K}$=A$_{K}$/E(H-K)) is different. We have used the values of \citet{rieke85} (R$_{K}$=1.78), \citet{espinoza09} (R$_{K}$=1.61) and \citet{nishi09} (R$_{K}$=1.44) to obtain the K-band extinction A$_{K}$. For the same observed color (H-K), the extinction can differ by 0.8 magnitude depending on the extinction law. The mass of star F12 varies between 93 and 138 \msun (35\% variation). In the bottom panel of Fig.\ \ref{fig_arches}, we show the most extreme variation in luminosity expected for the Arches star. For star F12, the initial mass can be as low as 78 \msun or as large as 146~\msun\ depending on the combination of photometry, extinction and distance adopted\footnote{For masses above 120 \msun, we simply linearly extrapolate the luminosities and initial masses from the grid of \citet{mm03}, taking the 85 and 120 \msun tracks as reference to estimate the dependence of mass on luminosity.}. There is almost a factor of two in the estimated initial mass.

From the bottom panel of Fig.\ \ref{fig_arches} we can conclude that according to the evolutionary tracks of \citet{mm03}, the most massive stars in the Arches cluster likely have masses higher than 80-90 \msun. Under the most favourable assumptions, the most massive objects can reach 150 to 200 \msun. This is the initial masses of stars with \lL\ $>$ 6.5 (the group of stars around \lL=6.55 corresponds to initial masses of 160 \msun). The truth probably lies somewhere in between these extreme values. 

The above mass estimates have been obtained using the evolutionary tracks of \citet{mm03}. The grid of \citet{ek12} is a revised version of these tracks, with a lower metal content (Z=0.014 versus 0.02). \citet{yusof13} also published models for very massive stars at solar metallicity (Z=0.014). Interestingly, none of the evolutionary calculations of Ekstroem et al.\ nor of Yusof et al.\ is able to reproduce the luminosity and temperature of the Arches stars. Their tracks for masses above 100 \msun\ remain close to the zero age main sequence or even turn rapidly towards the blue part of the HR diagram. They never reach temperatures lower than about 40000 K. \citet{paul10} pointed out that metallicity in the Galactic center may be slightly super solar \citep{paco09}. Evolutionary models with Z=0.02 might be more relevant for the Arches cluster. In any case, the behaviour of evolutionary tracks at very high masses indicates that large uncertainties exist in the predicted temperature and luminosity of very massive stars. This should be kept in mind when quoting initial masses derived by means of evolutionary calculations.

%
\subsubsection{R136 in 30~Doradus}
\label{s_arches}

R136 was once thought to be a single object of mass $\sim$ 1000 \msun, as described in Sect.\ \ref{hist}. As shown in Fig.\ \ref{fig1}, three components are resolved by adaptive optics observations. Two other bright objects (R136b and R136c) are located next to the R136a stars. Contrary to the case of the stars in the Arches cluster, extinction is not the main limitation of the luminosity and mass determination. Crowding is at least as important. It is only when the \textit{Hubble Space Telescope} started its operation that spectroscopic data could be obtained. Even with HST, a1 and a2 are difficult to separate. 

\begin{figure}[t]
     \centering
     \subfigure[]{ \includegraphics[width=.47\textwidth]{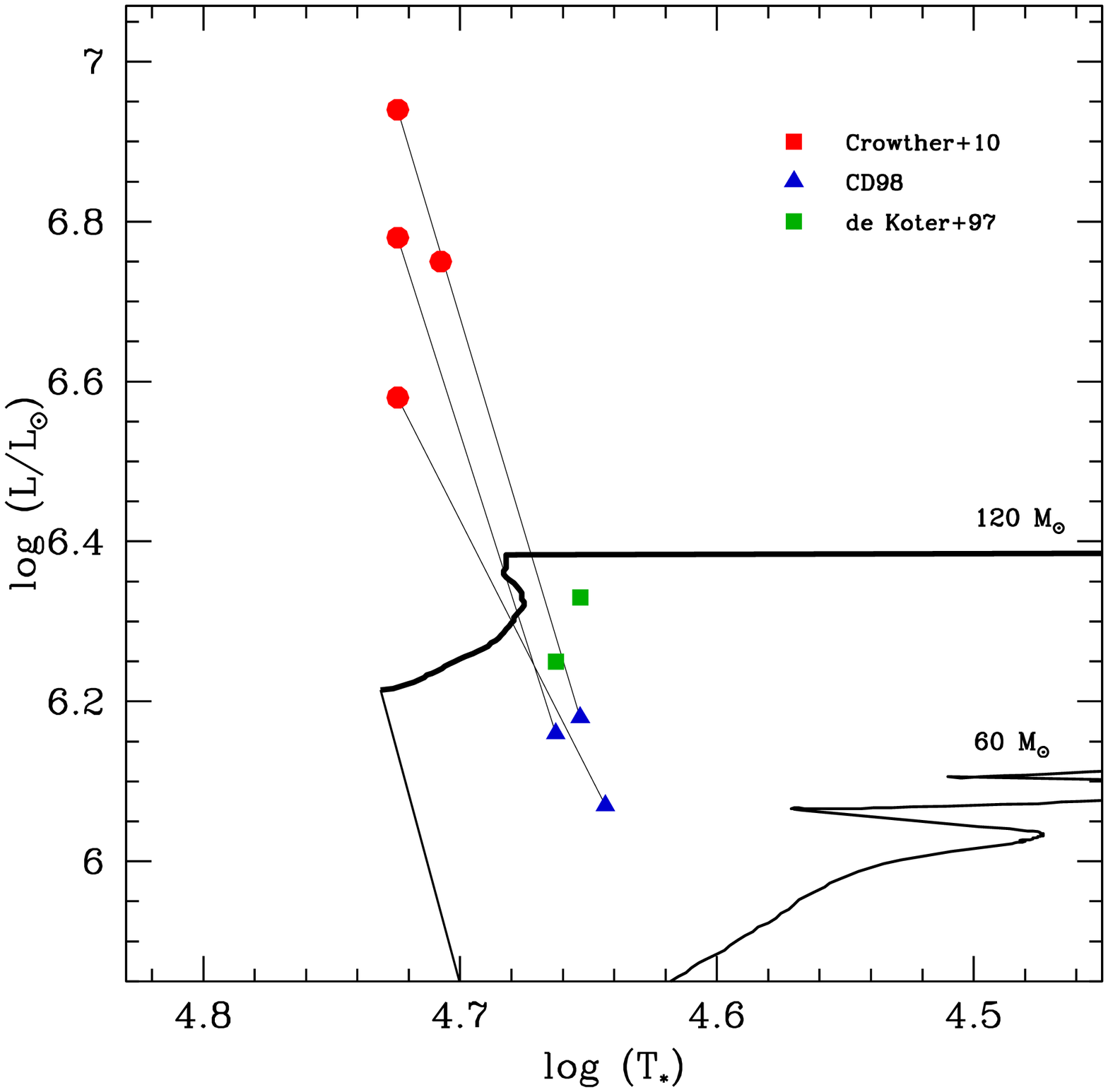}}
     \hspace{0.1cm}
     \subfigure[]{\includegraphics[width=.47\textwidth]{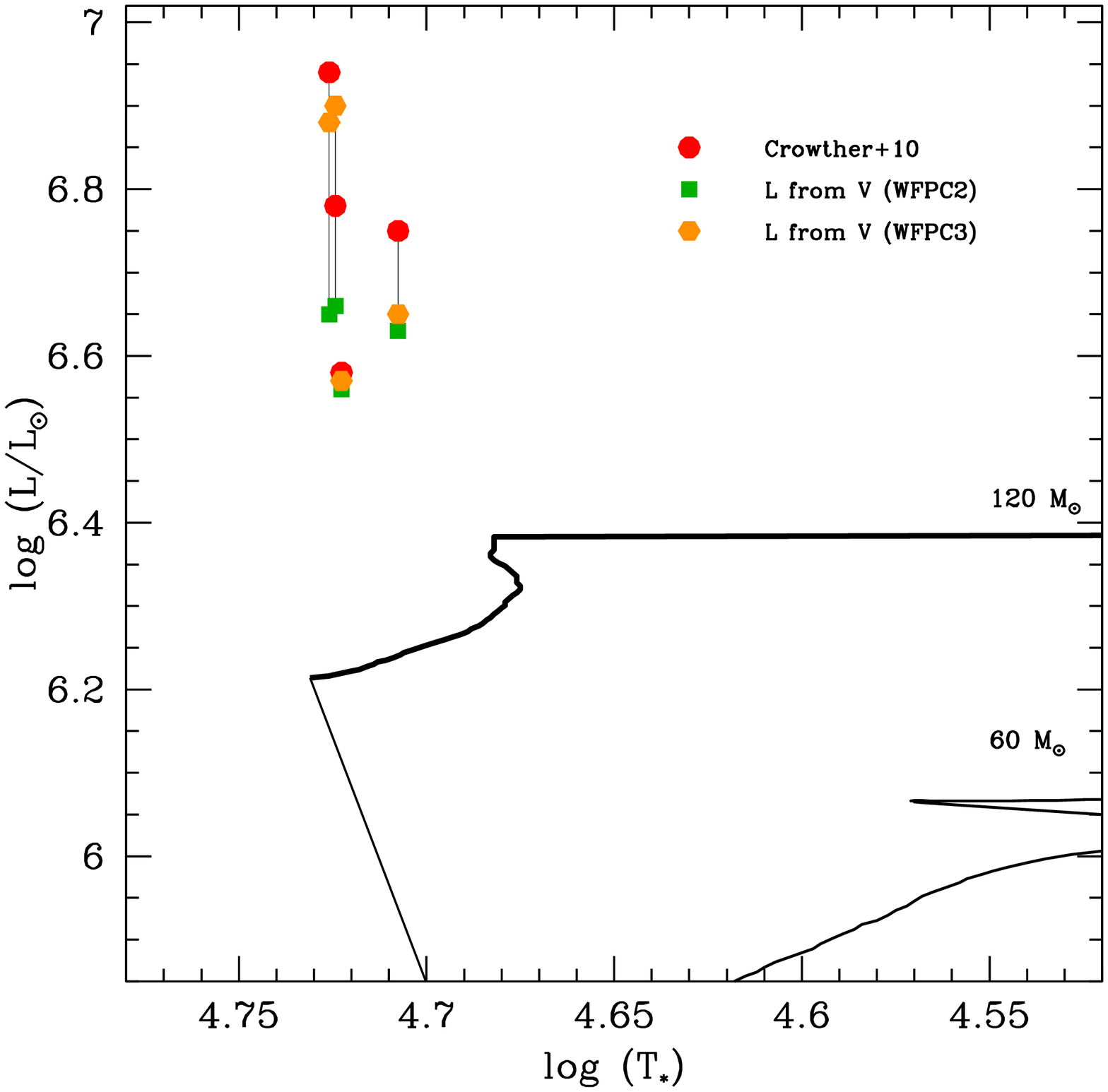}}
\caption{Hertzsprung-Russell for the stars in R136. The evolutionary tracks at Z=0.008 are from \citet{mm05}. The positions of the stars a1, a2, a3 and c are shown in red according to the analysis of \citet{paul10}. \textit{Left}: the green squares are stars a1 and a3 from \citet{dek97}. The blue triangles are stars a1, a2 and a3 from \citet{cd98}. \textit{Right}: the green squares corresponds to luminosity estimates based on HST/WFPC2 photometry \citep{hunter95}, effective temperatures and extinction from \citet{paul10}, and bolometric corrections from \citet{mp06}. The orange hexagons are based on HST/WFPC3 photometry from \citet{demarchi11}. The effective temperatures of stars a1, a2 and a3 have been shifted by 200 K for clarity.}
\label{hr_r136}
\end{figure}

\citet{dek97} performed an analysis of the UV spectra of a1 and a3. Their results are shown by the green points in Fig.\ \ref{hr_r136}. They determined effective temperatures of about 45000 K and luminosities close to 2.0 $\times 10^6$ L$_{\odot}$, corresponding to initial masses between 100 and 120 \msun. \citet{cd98} used improved model atmospheres to re-analyze a1 and a3, and included also a2. They extended their study to the optical range. They found very similar effective temperatures and luminosities 0.1 to 0.2 dex lower. A major re-investigation of the properties of the R136a components was performed by \citet{paul10}. They used atmosphere models including many metals, thus correctly taking into account the line-blanketing effects. In addition, they used the K-band spectra and photometry obtained by \citet{schnurr09}, taking advantage of the high spatial resolution collected with the adaptive optics demonstrator \textit{MAD} on the VLT \citep{campbell10}. Due to the inclusion of line-blanketing in atmosphere models, new, hotter effective temperatures were obtained (\teff\ = 53000 K). The luminosities were significantly revised upward, reaching 10$^{6.6-6.9}$ L$_{\odot}$ (see Fig.\ \ref{hr_r136}, left panel). Consequently, much larger initial masses were estimated: 320 \msun\ for a1, 240 \msun for a2, 165 \msun\ for a3 and 220 \msun\ for c.

\begin{table}[t]
\begin{center}
\caption{Luminosity estimates for the R136a stars using WFPC2 and WFPC3 photometry$^a$.}
\label{tab_r136}       
%
%
\begin{tabular}{p{1.3cm}p{1.3cm}p{2.1cm}p{1.3cm}p{2.1cm}p{1.3cm}p{1.1cm}}
\hline\noalign{\smallskip}
Star      &  \teff\  &  mV(WFPC2) &  A$_{V}$  &  MV(WFPC2) & BC(V)  &  \lL   \\
          &  [K]     &  mV(WFPC3) &      &  MV(WFPC3) &        &    \\
\noalign{\smallskip}\svhline\noalign{\smallskip}
R136a1   &  53000   &  12.84     & 1.80 &  -7.41 & -4.48 & 6.65  \\
          &          &  12.28     &      &  -7.97 &       & 6.88  \\
R136a2   &  53000   &  12.96     & 1.92 &  -7.41 & -4.48 & 6.66 \\
          &          &  12.34     &      &  -8.03 &       & 6.90  \\
R136a3   &  53000   &  13.01     & 1.72 &  -7.16 & -4.48 & 6.56  \\
          &          &  12.97     &      &  -7.20 &       & 6.57  \\
R136c    &  51000   &  13.47     & 2.48 &  -7.46 & -4.37 & 6.63  \\
          &          &  13.43     &      &  -7.50 &       & 6.65  \\
\noalign{\smallskip}\hline\noalign{\smallskip}
\end{tabular}
\end{center}
$^a$ A distance modulus of 18.45 is adopted. Bolometric corrections are from \citet{mp06}. Photometry is from \citet{hunter95} for HST/WPFC2 and \citet{demarchi11} for HST/WPFC3. Extinction is from \citet{paul10}.
\end{table}

There are several reasons for the revised luminosities: change of effective temperature and new bolometric corrections, different extinction, and use of different wavelength bands. \citet{dek97} and to a lower extent \citet{cd98} used atmosphere models with a limited treatment of line-blanketing. \citet{paul10} included many metals in their models. As is well known, metals affect the determination of effective temperatures and bolometric corrections \citep{msh05}. Let us take the example of star R136a3. \citet{dek97} used an absolute visual magnitude MV=-6.73 and obtained \teff\ = 45000 K. According to the calibration of \citet{vacca96} based on models without line-blanketing, the corresponding bolometric correction is -4.17. Using Eq.\ \ref{eq5} \citep[for the optical range, see][]{mp06} we obtain \lL\ = 6.26, in perfect agreement with the value of 6.25 determined by \citet{dek97}. If we were using the lower effective temperature of \citet{cd98} - \teff\ = 40000 K - one would get \lL\ = 6.11, close to 6.07 as derived by \citet{cd98}. If we now adopt the effective temperature derived by \citet{paul10} (\teff\ = 53000 K) and use modern calibrations of bolometric corrections including the effects of line-blanketing \citep{mp06}, we obtain BC = -4.48 and \lL\ = 6.38. Hence, better atmosphere models and temperature determinations contribute to an increase of about 0.2-0.3 dex in luminosity. But compared to the value obtained by \citet{paul10} (\lL\ = 6.58) there is another 0.2 dex to explain. \citet{dek97} adopted an extinction A$_{V}$=1.15, using E(B-V) = 0.35 and a standard value of the ratio of selective to total extinction. \citet{paul10} used the extinction determined by \citet{fs84} and obtained A$_{V}$ = 1.72 for star a3. If we use this extinction, the absolute magnitude reaches -7.24 which, together with BC = -4.8 leads to \lL\ = 6.59, similar to \citet{paul10}. For star a3, the combination of new effective temperature, bolometric corrections, and a larger extinction explains the revised luminosity and thus initial mass. The results are summarized in Table \ref{tab_r136} and Fig.\ \ref{hr_r136} (right panel). Table \ref{tab_r136} includes the optical photometry of \citet{hunter95} (HST/WFPC2) and that of \citet{demarchi11} (HST/WFPC3). Both are similar for star a3. 

If we now turn to star R136c, the new stellar parameters together with optical HST photometry lead to \lL\ = 6.65. This is lower than the value of \citet{paul10} by 0.1 dex, but still within the error bars. We also note that the WFPC2 and WFPC3 results are consistent. This is not true for stars a1 and a2. Using the WFPC2 magnitudes, we obtain \lL\ = 6.65 (6.66) for a1 (a2), while using WFPC3 photometry, one gets \lL\ = 6.88 (6.90) for a1 (a2). This values have to be compared to \lL\ = 6.94 and 6.78 obtained by \citet{paul10}. For both stars, WFPC2 photometry leads to lower luminosities than the K-band analysis of \citet{paul10}. The difference is significant for star a1. Luminosities are much higher if recent optical photometry is used. The results are now consistent for a1 (optical / K-band) while the optically derived luminosity of a2 is higher than the K-band luminosity. The changes in optical photometry might be due to crowding, stars a1 and a2 being separated by only 0.1''.

\begin{table}[t]
\begin{center}
\caption{Initial mass estimates for the R136a stars. M$_{C10}^{initial}$ and M$_{C10}^{current}$ are initial and current masses obtained from evolutionary tracks by \citet{paul10}. M$_{G11}$ are upper limits from the homogeneous M-L relations of \citet{graef11} using a hydrogen mass fraction of 0.7 and the luminosities based on K-band, WFPC2 and WFPC3 photometry (see Table \ref{tab_r136}), assuming the distance and extinction of \citet{paul10}. All values are in M$_{\odot}$.}
\label{tab_M_r136}  
\begin{tabular}{p{1.3cm}p{1.3cm}p{1.3cm}p{1.3cm}p{1.3cm}p{1.3cm}}
\hline\noalign{\smallskip}
Star      & M$_{C10}^{initial}$  & M$_{C10}^{current}$ & M$_{G11}^{K}$  & M$_{G11}^{WFPC2}$  & M$_{G11}^{WFPC3}$ \\
\noalign{\smallskip}\svhline\noalign{\smallskip}
R136a1   & 320$^{+100}_{-40}$ & 265$^{+80}_{-35}$ & 372 & 230 & 336 \\
R136a2   & 240$^{+45}_{-45}$  & 195$^{+35}_{-35}$ & 285 & 234 & 347 \\
R136a3   & 165$^{+30}_{-30}$  & 135$^{+25}_{-20}$ & 206 & 200 & 203 \\
R136c    & 220$^{+55}_{-45}$  & 175$^{+40}_{-35}$ & 271 & 223 & 231 \\
\noalign{\smallskip}\hline\noalign{\smallskip}
\end{tabular}
\end{center}
\end{table}

In Table \ref{tab_M_r136} we have gathered different mass estimates. The initial and current masses of \citet{paul10} are shown in the first two columns. They are based on the dedicated evolutionary tracks presented in the study of Crowther et al. Masses can also be obtained from mass-luminosity relations. \citet{graef11} presented such relations for completely mixed stars. In that case, there is no gradient of molecular weight inside the star and the mass is maximum for a given luminosity. We have used their relation for a hydrogen mass fraction of 0.7 (slightly higher than the observed values) to provide \textit{upper limits} on the current masses of the R136 stars. We have used the luminosities of \citet{paul10} and those given in Table \ref{tab_r136}, based on WFPC2 and WPFC3 photometry. The results are shown in the last three columns of Table \ref{tab_M_r136}. The upper limits are consistent with the current mass estimates of \citet{paul10}. The impact of luminosity on the derived mass is clearly visible from Table \ref{tab_M_r136}: R136~a1 has an upper mass limit between 230 and 372 \msun depending on its luminosity estimate.

\vspace{0.5cm}

The main conclusions regarding the revised R136 luminosities (and consequently masses) can be summarized as follows: 1) line-blanketed models lead to new effective temperatures and bolometric corrections which in turn contribute to an increase in luminosity; 2) higher visual extinction provides an additional source of luminosity increase; 3) photometric precision is important, especially in crowded regions. These three factors all affect the determination of luminosities, and thus initial masses, for stars in R136. Future studies of this region with high spatial resolution will certainly shed new light on the initial mass and nature of the R136 very massive components.

%
\subsubsection{NGC~3603}
\label{s_3603}

NGC~3603 is a Galactic massive cluster dominated by three bright objects named A, B and C. The former is a known binary and will be described in Sect.\ \ref{vms_bin}. Stars B and C have been analyzed by \citet{cd98} and revisited by \citet{paul10}. They are Wolf-Rayet stars of type WN6h. 

\citet{cd98} obtained \lL\ = 6.16 and 6.06 for stars B and C respectively. They did not use luminosities to infer stellar masses, but instead relied on the wind properties \citep{kud92}. According to the radiatively driven wind theory, the wind terminal velocity is directly related to the escape velocity, which is itself related to the surface gravity. With the effective temperature and luminosity obtained from the spectroscopic analysis, the radius is known. Calculations of the terminal velocity in the framework of the radiatively driven wind theory and comparison to the observed velocities provides an estimate of the present stellar mass. \citet{cd98} obtained M = 89$\pm$12 (62$\pm$8) for star B (C).

\citet{paul10} used the more classical transformation of luminosities into stellar masses in the HR diagram. As for R136, they used atmosphere models including a proper treatment of line-blanketing effects. Consequently, the effective temperatures they derive are higher. They also take into account a larger extinction (A$_{V}$ $\sim$ 4.7 versus 3.8 for Crowther \& Dessart) and a shorter distance (7.6 kpc versus 10.0 kpc, corresponding to distances modulus of 14.4 and 15.0 respectively). The effect of extinction and distance act in different directions, but the former is larger, so that in addition to the increased luminosity due to higher \teff, the luminosities obtained by \citet{paul10} are higher than \citet{cd98}: \lL\ = 6.46 and 6.35 for B and C respectively. The present day masses (113 \msun\ for both stars) and initial masses (166$\pm$20 and 137$^{+17}_{-14}$ \msun) are higher than the wind masses quoted above. They were obtained using non-rotating tracks. Since evolutionary models including rotation have higher luminosities, lower initial masses would be determined if they were used.

The determination of the masses of the brightest objects in NGC~3603 illustrates once again the role of extinction and of better stellar parameters. But it also highlights the importance of distances. For the Arches cluster and R136, they were quite well constrained. For NGC~3603, the uncertainty is larger.  Assuming the physical parameters of \citet{paul10} for star B, but using a distance of 10.0 kpc (instead of 7.6) would lead to \lL\ = 6.70, 0.24 dex higher than reported by Crowther et al. The initial masses would then exceed 200 \msun. Accurate parallaxes hopefully provided by the \textit{Gaia} mission will help to refine the mass estimates of NGC3603 B and C.

%
\subsubsection{Other candidates}
\label{s_other}

There are a few stars that can be considered as candidates to have masses in excess of 100 \msun. Most of them are located in the Galactic Center. The first one is the Pistol star \citep{figer98}. As we have described in Sect.\ \ref{hist}, it was once thought to have a luminosity between 1.5 and 4.0 $\times$ 10$^6$ L$_{\odot}$, corresponding to a mass larger than 200 \msun. But the luminosity was revised by \citet{paco09} so that the current initial mass of Pistol is thought to be closer to 100 \msun. In addition, \citet{mart11} reported the presence of a faint companion, calling for an additional downward revision of the mass of the brightest member.

If the Pistol star has lost its status of most luminous star in the Galactic Center, other objects have attracted attention in the last years. \citet{barniske08} studied two WN stars located between the Arches and central cluster, in relative isolation. Using 2MASS near-infrared and Spitzer mid-infrared photometry, they fitted the SED of these stars and determined \lL\ = 6.3$\pm$0.3 and 6.5$\pm$0.2 for stars WR102c and WR102ka respectively. They adopted the extinction law of \citet{rieke85} and \citet{lutz99}, as compiled by \citet{moneti01}. Keeping in mind the limitations described in Sect.\ \ref{s_arches}, these two stars are comparable to the most luminous WNh stars in the Arches cluster, and could have initial masses between 100 and 150 \msun. 

\citet{hamann06} studied most of the Galactic WN stars and obtained luminosities larger than 2.0 $\times$ 10$^6$ L$_{\odot}$ for a few objects (WR24, WR82, WR85, WR131, WR147, WR158). All are H-rich WN stars, like all luminous stars in young massive clusters presented above. This seems to be a common characteristics of all suspected very massive stars: they have the appearance of WN stars, but contain a large hydrogen fraction. They are thus most likely core-H burning objects, still on the main sequence. Their Wolf-Rayet appearance is due to their hight luminosity and consequently their strong winds.  

Finally, \citet{besten11} reported on VFTS-682, a H-rich WN star located 29 pc away (in projection) from R136. Bestenlehner et al.\ rely in optical spectroscopy and optical-near/mid infrared photometry to constrain the stellar parameters. They determine an effective temperature of 54500$\pm$3000 K. The luminosity estimate depends on the assumptions made regarding the extinction law. With a standard R$_V$=3.1, they obtain \lL\ = 5.7$\pm$0.2. But the resulting SED does not match the 3.6 $\mu$m and 4.5 $\mu$m \textit{Spitzer} photometry. To do so, the authors use a modified extinction law with R$_V$=4.7. The SED is better reproduced and the luminosity increases to \lL\ = 6.5$\pm$0.2. The \textit{Spitzer} 5.8 $\mu$m and 8.0 $\mu$m photometry remains unfitted, which is attributed to mid infrared excess possibly caused by circumstellar material. Bestenlehner et al. also argue that differences in the near infrared measurements between IRSF and 2MASS translates into an additional uncertainty of 0.1 dex on the luminosity. They favour the highest luminosity, which would correspond to a mass of about 150 \msun. A luminosity of 10$^{5.7}$ L$_{\odot}$ would correspond to an initial mass of about 40 \msun (see Sect.\ \ref{s_L}).


\vspace{0.5cm}

In conclusion, there are several stars that can be considered good candidates for a VMS status. The best cases are located in the Aches, NGC~3603 and R136 clusters. However, we have highlighted the uncertainties affecting their luminosity determination and consequently their evolutionary masses. If masses in excess of 100 \msun\ are likely, values higher then 200 \msun\ are still subject to discussion and should be confirmed by new analysis.

%
\section{Very Massive Stars in binary systems}
\label{vms_bin}

In this section we focus on binary systems containing massive stars. After presenting the various types of binary stars, we recall some relations relating physical to observed properties of binary systems. Finally, we present the best cases for binary systems containing very massive stars. 

%
\subsection{Massive binaries and dynamical masses}
\label{s_orbits}

%
\subsubsection{Types of binaries}
\label{s_types}

\citet{mason98} showed that two main types of binaries are detected: visual binaries and spectroscopic systems. The former are seen in imaging surveys. The two components of the systems are observed and astrometric studies covering several epochs can provide the orbital parameters, especially the separation and period. Visual binaries have periods larger than 100 years. The majority actually have periods of 10$^4$ to 10$^6$ years (corresponding to separations of the order of 0.1 to 10 pc). The reason is that massive binaries are located at large distances. Even with the best spatial resolution achievable today, only the systems with the widest separation can be detected. The other category of binaries is observed by spectroscopy. Due to the orbital motion of the two components around the center of mass, spectral lines are Doppler shifted. Radial velocities can be measured. If a periodicity is found in the RV curve, this is a strong indication of binarity. Spectroscopic binaries have short periods: from about 1 day to a few years, with a peak between 3 days and 1 month (separations of the order of a few tenths to one AU). This is again an observational bias: to detect significant variations of radial velocities the components have to be relatively close to each other, which implies short periods. Spectroscopic binaries cannot be spatially resolved. Hence, the spectrum collected by observations is a composite of the spectra of both components. Depending on the line strengths and the luminosity of the components, the spectral signatures of only one star or of both components can be observed. In the former (latter) case, the system is classified as SB1 (SB2). Spectroscopic binaries can sometimes experience eclipses if the inclination of the system is favourable. In that situation, the components periodically pass in front of each other, creating dips in the light curve. Such systems are the best to constrain masses as we will see below. Binary stars with periods between a few years and a century currently escape detection. 

In the following we focus on spectroscopic binaries since all the suspected very massive stars in multiple systems discovered so far belong to this category.

%
\subsubsection{Orbital elements and dynamical masses}
\label{s_orbits}

The first step to study a spectroscopic binary is to measure the radial velocities from spectral lines. The most widely used method consists in fitting Gaussian profiles to the observed lines. Such profile are not able to account for the real shape of O stars lines, but it is a good approximation for their core, at least in the optical range. This is sufficient to measure a radial velocity (RV). Several lines are usually used and the final RV is the average of the individual RV measured on each line. An important limitation of this method is that it assumes that lines are formed in the photosphere at zero velocity with respect to the star's center of mass. For absorption lines of O stars, this is a good approximation. For stars with strong winds such as WNh stars, the observed lines will be mainly formed in the wind. Hence, they will have a non zero velocity due to the wind outflow. This can introduce a systematic offset in the radial velocity curve. Hopefully, the important quantity to constrain the mass of binary components is the amplitude of the RV curve, which is less sensitive to the above limitation. A way to avoid this problem is to use a cross-correlation method in which the spectrum of a single star with a spectral type similar to the components of the system is used to obtain RV. 

The vast majority of spectroscopic binaries are studied using optical spectra. Hydrogen and helium lines are the main indicators, especially when the Gaussian fit method is employed. For the very massive stars we will describe below, nitrogen lines are also present and are included in the analysis. Near-infrared spectra are becoming available and are well suited to investigate the binarity of stars hidden behind several magnitudes of extinction. The K-band is the most commonly used spectral range in the near-infrared. Here again, hydrogen and helium lines are present. The typical width of absorption lines in O stars is 10 to 20 km s$^{-1}$. In order to correctly resolve them, at least ten points should be obtained throughout the line profile. This implies a minimum spectral resolution of $\sim$ 3000. The higher the spectral resolution, the smaller the error on the RV measurement. For stars with strong winds, lines are broad and high spectral resolution is not required.

\begin{figure}[t]
\begin{center}
\includegraphics[width=7cm]{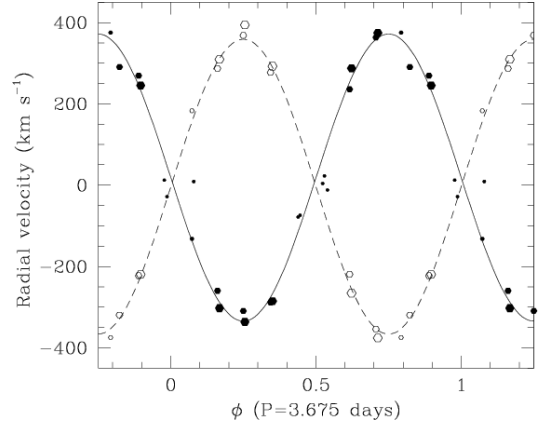}
\caption{Radial velocity curve of the binary system WR20a. Filled (open) symbols are for the primary (secondary). From \citet{rauw04}. Reproduced with permission.}
\label{rv_wr20a}
\end{center}
\end{figure}

Once the radial velocity measurements are obtained, a period search can be performed using time series analysis. If a signal is detected, radial velocities can be phased to create the type of curves presented in Fig.\ \ref{rv_wr20a}. In a SB2 system, the RV changes of both components are seen. They are anti-correlated. In a SB1 system, only the radial velocity variations of the brightest star are seen. According to Kepler's laws, the semi-amplitude of the RV variations (K) is related to the orbital elements as follows:

\begin{equation}
K = (2 \pi\ G)^{1/3} (\frac{M}{T})^{1/3} \frac{q}{(1+q)^{2/3}} \frac{\sin i}{\sqrt{1-e^2}}
\label{rel_rv}
\end{equation}

\noindent where $G$ is the constant of gravitation, $M$ the mass of the star, $T$ the period, $q=M_{2}/M_{1}$ the ratio of the secondary to primary mass, $i$ the inclination of the system and $e$ its eccentricity. For a SB1 system, we can reprocess this relation to obtain the so-called mass function

\begin{equation}
f(m) = \frac{M \sin^3 i}{(1+q)^2} = \frac{K^3}{2 \pi G} (1-e^2)^{3/2} T
\end{equation}

\noindent This is not sufficient to constrain the mass of the system unless assumptions on the mass ratio are made. On the contrary, for a SB2 system we can use Kepler first law to derive

\begin{eqnarray}
M_{1} \sin^3 i = \frac{T}{2 \pi G} (K_{1}+K_{2})K_{2} (1-e^2)^{3/2}\\
M_{2} \sin^3 i = \frac{T}{2 \pi G} (K_{1}+K_{2})K_{1} (1-e^2)^{3/2}
\label{eq_rv}
\end{eqnarray}

\noindent from which we immediately see that the mass ratio $M_{1}/M_{2}$ is the inverse of the RV amplitude ratio $K_{2}/K_{1}$.

With these expressions, it is possible to fit the observed RV curves and to estimate lower limits on the components masses. These masses are called ``dynamical masses''. The shape of the RV curve provides a first guess of the eccentricity. For a circular orbit, it should be a perfect sine curve. For more eccentric orbits, the RV curve becomes asymmetric with a maximum absolute value of the radial velocity concentrated in a short period of time corresponding to periastron passage. In Fig.\ \ref{rv_wr20a} the observations are consistent with a zero eccentricity. The ratio of the semi-amplitude K$_{1}$/K$_{2}$ is the inverse of the mass ratio. Consequently, a look at the RV curve already provides information on the mass ratio of the system: for equal mass binaries, the amplitude of the radial velocity variations of the primary and of the secondary stars are the same. From the above equations, it appears that the determination of M$\sin^3 i$ is essentially model independent. The main limitations come from the measurement of the radial velocities.

To obtain dynamical masses and not just lower limits, one needs to determine the inclination of the system. This is possible for eclipsing binaries. A photometric monitoring over several periods provides the light curve, i.e. the evolution of the system's magnitude as a function of time. For eclipsing binaries, minima are observed in the light curve. They correspond to the physical situation where one star passes in front of the other and (partly) blocks its light. The shape of the light curve in the eclipse phases can be quite complex depending on the geometry of the system. For stars relatively separated and experiencing complete eclipses, a plateau is observed in the minimum of the light curve. If the eclipse is only partial, no plateau is seen. For close binaries, interactions between components affect the shape of the stars and their light. Tidal forces cause elongations along the binary axis. Light of one component can be reflected and/or heat the surface of the companion, making more complex the shape of the eclipses in the light curve. In the most extreme cases of contact binaries, material is exchanged between both stars which affects the system's photometry. All these processes render the fit of the light curve more difficult and more mode-dependent than the RV curve. Even in the simplest case of detached, non-interacting binaries, a limb darkening model has to be adopted to reproduce the eclipses. In addition, assumptions on the luminosity ratio are usually necessary to obtain a complete solution. Under these assumptions, it is possible to constrain the inclination of the system and thus, from Eq.\ \ref{eq_rv}, the dynamical masses.

In order to refine the analysis of RV and light curves, a better knowledge of the physical properties of the stars is useful. The method of spectral disentangling consists in the separation of the individual spectra from the combined, observed spectrum of the system. The principle is rather simple: subtract the spectrum of one star from the combined spectrum. In practice, this method requires again a good knowledge of the luminosity ratio of both components in order to evaluate their respective contribution. The radial velocity amplitudes have to be large enough so that the two spectra are well separated. A good sampling of the orbit is necessary. Under these conditions, it is possible to extract the individual spectra and perform spectral classification. A spectroscopic analysis of each star can be attempted, providing additional stellar parameters. Like for single stars, evolutionary and spectroscopic masses can be obtained and directly compared to dynamical masses \citep[e.g.][]{mahy11}.

Provided that eclipses are observed, it is thus possible to obtain dynamical masses for massive binaries. The analysis of the RV curve is affected mainly by uncertainties in the measurements of radial velocities, while the interpretation of the light curve is much more model dependent. With these limitations in mind, we now present a few example of massive binary systems potentially hosting VMS.

%
\subsection{The most massive binary systems}
\label{s_candidates_bin}

Now that the basic ingredients of mass determination in binary systems have been recalled, we turn to the presentation of some of the most interesting systems potentially hosting very massive stars. 

%
\subsubsection{WR20a}
\label{s_wr20}

WR20a is located in the vicinity of the massive cluster Westerlund~2. \citet{rauw04} presented the optical spectrum of the system and first radial velocity measurements. They used Gaussians to fit the main emission lines. The main difficulty they encountered was that the two components have the same spectral type: WN6ha. Hence, a period search was performed in the absolute difference of radial velocities ($\mid RV_{1}-RV_{2} \mid$). The preferred value of the period was 3.675 d, with a second possible value at 4.419 d. Assuming a zero eccentricity orbit (from the shape of the RV curve), \citet{rauw04} determined minimum masses of 70.7$\pm$4.0 and 68.8$\pm$3.8 \msun\ for the primary and secondary. 

At the same time, \citet{bonanos04} collected I-band photometry for WR20a. The lightcurve showed clear minima, revealing the eclipsing nature of the system. They revised the period determination: 3.686 d. Performing a fit of the light curve, \citet{bonanos04} determined an inclination of 74$^o$.5$\pm$2$^o$.0. Combined with the radial velocity measurements of \citet{rauw04}, Bonanos et al.\ concluded that the current masses of the two components are 83.0$\pm$5.0 and 82.0$\pm$5.0 \msun. The very similar masses are consistent with both stars having the same spectral type, and with the absence of strong asymmetry in the light curve (all eclipses have almost the same depth). \citet{rauw05} subsequently determined the stellar and wind parameters of the system and obtained an effective temperature of 43000 K and a luminosity of 1.15 $\times$ 10$^6$ L$_{\odot}$ for each component. The effective temperature is consistent with that assumed by \citet{bonanos04} for their light curve fit. Using the evolutionary tracks of \citet{mm03}, the stellar parameters corresponds to  present mass of 71 \msun\ and an initial mass of 84 \msun. Given the uncertainties related to evolutionary tracks described in Sect.\ \ref{s_tracks}, the agreement is good. 

Although not a very massive star per see, WR20a is one of the binary star with the best determination of dynamical masses.

%
\subsubsection{NGC~3603 A1}
\label{s_A1}

In Sect.\ \ref{s_3603} we have shown that NGC~3603 contained two stars (B and C) with high luminosities and good candidates for having initial masses in excess of 100 \msun. A third object lies in the core of NGC~3603: star A1. Using unresolved spectroscopy, \citet{mn84} showed that object HD~97950, which includes all three stars A1, B and C, was variable in radial velocity. A period of 3.77 days could be identified. Subsequent investigation of NGC~3603 with HST by \citet{moffat04} revealed photometric variability with the same period. With the improved spatial resolution, the origin of the variability could be attributed to star A1 which was then classified as a double-line eclipsing binary (SB2). \citet{moffat04} performed a series of light curve fits assuming various values for the mass ratio of both components. They found that an inclination of 71$^o$ best reproduced their light curve. This inclination was obtained for a mass ratio between 0.5 and 2.0 and for an effective temperature of the secondary (primary) of 43000 K (46000 K).

\citet{schnurr08} used adaptive optics integral field spectroscopy in the K-band to monitor the radial velocity variations of A1. They showed that Br$\gamma$ and HeII 2.189 $\mu$m were both in emission and were double peaked, with one peak stronger than the other. They concluded that the system was most likely formed by a WN6ha star and a less luminous O star. They determined radial velocities from Gaussian fits to the emission lines. Measurements of the secondary radial velocity variations were difficult when lines from the primary and secondary were blended and dominated by the primary's emission. This resulted in large uncertainties on the secondary RV curve and consequently on the dynamical mass estimates. Schnurr et al.\ adopted the inclination and period of \citet{moffat04} as well as a zero eccentricity and calculated an orbital solution. They obtained M = 116$\pm$31 \msun\ and M = 89$\pm$16 \msun\ for the primary and secondary respectively. The mass ratio is thus 1.3$\pm$0.3. 

\citet{paul10} performed a spectroscopic study of the integrated spectrum of A1 at maximum separation of the two components. They found \teff\ = 42000~K for the primary and \teff\ = 40000 K for the secondary. This is lower than the values adopted by \citet{moffat04} for their light curve solution. Revisiting the inclination determination with these new temperatures would be useful. Crowther et al.\ also constrained the luminosity of the components and used non-rotating evolutionary tracks to estimate the current mass of the stars: 120$^{+26}_{-17}$ and 92$^{+16}_{-15}$ \msun. These values are in very good agreement with the dynamical masses. But here again, current masses determined with evolutionary tracks including rotation would have been lower.

%
\subsubsection{R144 and R145}
\label{s_r144}

R144 and R145 are two H-rich WN stars located in the LMC. \citet{moffat89} reported variability in R145 and tentatively derived a periodicity of 25.4 days, indicating a binary nature. \citet{schnurr09} monitored the spectroscopic variability of R145 and confirmed that most lines display changes in both shape and position. They observed mainly one set of line, thus one of the components. Its spectral is similar to other massive binaries: WN6h. Using a method based on the removal of the averaged primary spectrum in each individual spectra, they identified weak HeII features typical of O stars. Hence, they concluded that the secondary should be a less massive object than the primary. This was confirmed by estimates of the visible flux level and the study of the strength of the companion HeII lines in comparison with template spectra of O stars. The primary should be about 3 times more massive than the secondary.

In addition to spectroscopic observations, \citet{schnurr09} also obtained simultaneous polarimetric data in which they clearly detected Zeeman signatures in the Q and U profiles. Interestingly, these profiles also showed variability. A clear periodicity of 159 d could be identified. Using this period, they obtained an orbital solution for the radial velocity curve yielding M$sin^3i$ = 116$\pm$33 \msun (48$\pm$20 \msun) for the primary (secondary). The mass ratio of 2.4$\pm$1.2 was in reasonable agreement with that determined from the visible continuum flux. In a further step, \citet{schnurr09} performed a combined fit of the radial velocity and polarimetric curves and obtained $i$ = 38$^o\pm$9$^o$ (inclination of the system). With such an inclination, the primary and secondary mass should be larger than 300 and 125 \msun. These masses are inconsistent with the systems brightness though, questioning the inclination determination. 

R144 was also monitored by \citet{schnurr09}, but no periodicity was detected. \citet{sana13} presented new spectroscopic observations. They detected line shifts in NIII, NIV and NV lines. Interestingly, the NIII and NV shifts are anti-correlated, pointing towards a different origin for both types of lines. \citet{sana13} concluded that the optical spectra are the composite of the spectra of two types of stars: one WN5-6h mainly contributing to NV lines, and one WN6-7h star dominating the NIII features. A period search indicated a variability on a timescale of 2 to 12 months, without better constraint. Based on the visual photometry and assumptions regarding the bolometric correction of the components, \citet{sana13} estimated \lL\ $\sim$ 6.8. This places R144 among the potential very massive stars. 

For both R144 and R145, additional spectroscopic and photometric observations are required to better characterize the system's component. In particular, spectral disentangling would be helpful to constrain the nature of these massive systems.

\vspace{0.5cm}

In conclusion, there are relatively few binary systems potentially hosting VMS. The best cases would benefit for additional combined RV and light curve analysis (when eclipses are present) to refine the dynamical mass estimates. NGC~3603 A1 appears to be the best candidate.

%
\section{Summary and conclusions}
\label{s_conc}

In this chapter we have presented the observational evidence for the existence of very massive stars. Mass estimates of single stars are mainly obtained from the conversion of luminosities to evolutionary masses. We have highlighted the uncertainties in the determination of luminosities: crowding, accurate photometry, distance, extinction, atmosphere models all contribute to render uncertain luminosity estimates. The other source of error comes from evolutionary tracks. Different calculations produce different outputs depending on the assumptions they are built on. Even if the luminosity was perfectly well constrained, its transformation to masses relies on the predictions of evolutionary models.
With these limitations in mind, there are several stars that can be considered as good candidates for a VMS status. They are manily located in the massive young clusters NGC~3603, R136 and the Arches. In R136, the brightest members may reach initial masses higher than 200 \msun. In the other two clusters, masses between 150 and 200 \msun\ are not excluded. 
A few binary systems may also host stars with masses in excess of 100 \msun. NGC~3603 A1 seems to be the best candidate, with a M $\sim$ 115 \msun\ primary star, but a better analysis of the light curve is needed to refine the analysis. 

In conclusion, very massive stars may be present in our immediate vicinity. They usually look like WN5-9h stars, i.e. hydrogen rich mid to late WN stars. Super star clusters  -- the best places to look for VMS -- being impossible to resolve with the current generation of instruments, these local VMS have to be re-observed and re-analyzed in order to minimize the uncertainties involved in their mass determination. This is important to understand the upper end of the initial mass function and the formation process of massive stars in general.

%
%
\begin{acknowledgement}
The author thanks Paul Crowther for discussions on the mass determination of the R136 stars. 
\end{acknowledgement}

\bibliographystyle{aa}
\bibliography{martins_chapI.bib}

\end{document}